Novim

# Climate Engineering Responses to Climate Emergencies

**Jason J. Blackstock**[†]
**David S. Battisti**
**Ken Caldeira**
**Douglas M. Eardley**
**Jonathan I. Katz**
**David W. Keith**
**Aristides A. N. Patrinos**
**Daniel P. Schrag**
**Robert H. Socolow**
**and Steven E. Koonin**[†,‡]

[†] *Report Lead Authors*
[‡] *Study Group Convener*

**This report should be cited as either:**

J. J. Blackstock, D. S. Battisti, K. Caldeira, D. M. Eardley, J. I. Katz, D. W. Keith, A. A. N. Patrinos, D. P. Schrag, R. H. Socolow and S. E. Koonin, *Climate Engineering Responses to Climate Emergencies* (Novim, 2009), archived online at: http://arxiv.org/pdf/0907.5140

OR

J. J. Blackstock et al., *Climate Engineering Responses to Climate Emergencies* (Novim, 2009), archived online at: http://arxiv.org/pdf/0907.5140

# Climate Engineering Responses to Climate Emergencies

***Public Report Released:* July 28, 2009**

*Study Group Meeting:* August 10 to 15, 2008

***Study Group Participants:***

Jason J. Blackstock[†], David S. Battisti, Ken Caldeira, Douglas M. Eardley, Jonathan I. Katz, David W. Keith, Aristides A. N. Patrinos, Daniel P. Schrag, Robert H. Socolow and Steven E. Koonin[†,‡]

[†]*Report Lead Authors* [‡]*Study Group Convener*

## Table of Contents

### List of Figures

## Study Group Participants

| | |
|---|---|
| **Dr. Steven E. Koonin*** | Study Group Leader; Chief Scientist, BP; former Provost, Caltech |
| **Dr. Jason J. Blackstock** | Research Scholar, International Institute for Applied Systems Analysis (IIASA); Fellow, Centre for International Governance Innovation |
| **Dr. David S. Battisti** | Professor of Atmospheric Sciences, University of Washington |
| **Dr. Ken Caldeira** | Staff Scientist, Carnegie Institution for Science, Stanford University |
| **Dr. Doug M. Eardley** | Professor of Physics, Kavli Institute for Theoretical Physics, UCSB |
| **Dr. Jonathan I. Katz** | Professor of Physics, Washington University |
| **Dr. David W. Keith** | Director, ISEEE Energy and Environmental Systems Group, at the University of Calgary |
| **Dr. Aristides A. N. Patrinos** | Former Director of the Office of Biological and Environmental Research, DOE Office of Science; President of Synthetic Genomics |
| **Dr. Daniel P. Schrag** | Professor of Earth and Planetary Sciences, Harvard University; Director of the Harvard University Center for the Environment |
| **Dr. Robert H. Socolow** | Professor of Mechanical & Aerospace Engineering, Princeton Environmental Institute, Princeton University |

**Steven E. Koonin is currently the Under Secretary for Science at the US Department of Energy*

## Acknowledgements

The study participants and the Novim organization are grateful for the hospitality of the Kavli Institute for Theoretical Physics (KITP) at the University of California, Santa Barbara (UCSB) during the course of this study. Special thanks are extended to: Dr. David Gross (KITP Director); Michael Ditmore (Novim Executive Director); and Jim Knight (Novim Executive Vice President). The authors of this report are also grateful to our numerous colleagues who provided invaluable feedback on early drafts of this report.

## Executive Summary

Despite efforts to stabilize $CO_2$ concentrations, it is possible that the climate system could respond abruptly with catastrophic consequences. Intentional intervention in the climate system to avoid or ameliorate such consequences has been proposed as one possible response should such a scenario arise. In a one-week study, the authors of this report conducted a technical review and evaluation of proposed climate engineering concepts that might serve as a rapid palliative response to such climate emergency scenarios.

Because of their potential to induce a prompt (<1 yr) global cooling, this study concentrated on *Shortwave Climate Engineering* (SWCE) methods for moderately reducing the amount of shortwave solar radiation absorbed by the Earth. The study's main objective was to outline a decade-long agenda of technical research that would maximally reduce the uncertainty surrounding the benefits and risks associated with SWCE. For rigor of technical analysis, the study focused the research agenda on one particular SWCE concept—stratospheric aerosol injection—and in doing so developed several conceptual frameworks and methods valuable for assessing any SWCE proposal.

Basic physical science considerations, exploratory climate modeling, and the impacts of volcanic aerosols on climate all suggest that SWCE could partially compensate for some effects—particularly net global warming—of increased atmospheric $CO_2$. However, existing data also reveal important limits to the range of $CO_2$ impacts that SWCE could ameliorate; for example, ongoing ocean acidification would not be affected, and some categories of climate emergency scenario might prove unresponsive to SWCE. Moreover, significant uncertainty presently surrounds the spatial and temporal response of numerous climate and ecological parameters to SWCE, making the near-term deployment of large-scale SWCE extraordinarily risky.

Components of any comprehensive research agenda for reducing these uncertainties can be divided into three progressive phases: (*I*) *Non-Invasive Laboratory and Computational Research*; (*II*) *Field Experiments*; and (*III*) *Monitored Deployment*. Each phase involves distinct and escalating risks (both technical and socio-political), while simultaneously providing data of greater value for reducing uncertainties.

The core questions that need to be addressed can also be clustered into three streams of research: *Engineering* (intervention system development); *Climate Science* (modeling and experimentation to understand and anticipate impacts of the intervention); and *Climate Monitoring* (detecting and assessing the actual impacts, both anticipated and unanticipated). While a number of studies have suggested the engineering feasibility of specific SWCE proposals, the questions in the Climate Science and Climate Monitoring streams present far greater challenges due to the inherent complexity of temporal and spatial delays and feedbacks within the climate system.

These frameworks are applied to structure the comprehensive research agenda outlined for stratospheric aerosol SWCE in Part 3 of this report. For the Engineering stream, current understanding, questions and methods guiding the necessary research into aerosol material, stratospheric lofting and dispersion are all defined. For the Climate Science and Climate Monitoring streams, emphasis is placed on identifying, predicting and monitoring the response of important climate parameters across four broad categories: *Radiative*, *Geophysical*, *Geochemical* and *Ecological*. Finally, the components within each stream are identified as belonging to *Phase I* or *II* research, and the limits placed by the natural variability of the climate system on what can be learned from low-level *Phase II* field-testing are roughly assessed.

This report does not attempt to evaluate whether stratospheric aerosol (or any other) SWCE systems *should* be developed or deployed—or even whether any parts of the outlined research program should be pursued. Such questions are the subject of an intense ongoing debate, involving socio-political and economic issues beyond the scope of this study. This report aims to better inform that debate by elucidating the technical research agenda that would be necessary to reduce the uncertainty in potential SWCE interventions.

# PRELUDE

## THE CONTEXT OF THE "SHOULD" DEBATE ABOUT CLIMATE ENGINEERING RESEARCH

The limited scientific research conducted to date on climate engineering is not accidental. As articulated by Cicerone (2006), such ideas have "not enjoy[ed] broad support from scientists."[1] This circumstance stems principally from four core concerns shared to varying degrees by many scientists:[2]

(1) The climate system might be inherently too complex—and therefore the possibility of "[unanticipated] harmful side effects"[3] too large—for any intentional human intervention to ever be considered safe.

(2) Shortwave climate engineering can be perceived as a substitute for greenhouse gas (GHG) emission reductions, and might therefore "undercut human resolve to deal with the cause of the original problem."[4]

(3) If a widespread political belief developed that climate control is (or will become) possible through climate engineering, significant international tensions might emerge surrounding who gets to define what the "optimum" climate should be.[5]

(4) There are a host of ethical beliefs about humanity's role in the natural world that are opposed to intentional human modification of the Earth's climate.

Arguments about the validity and importance of these concerns are at the heart of the ongoing debate about whether comprehensive climate engineering research *should* be conducted. As several of these concerns have clear economic (2), socio-political (2 & 3) or ethical (4) roots, they were largely outside the expertise of the study group participants and thus beyond the scope of this study.

For that reason, this report does not address whether (or which parts of) the uncertainty-reducing research agenda *should* be pursued. It therefore should not be interpreted as advocacy for a particular research agenda or as an attempt to pre-empt the ongoing debate in the mainstream scientific and public arenas. There are almost certainly tradeoffs in the socio-political dimensions between reducing uncertainty in possible climate interventions and the socio-political risks created by pursuing specific components of the research agenda.

---

[1] Cicerone-2006.

[2] Readers can find expanded discussion of these issues in a handful of editorial essays, including Cicerone's, that accompanied Paul Crutzen's 2006 paper (Crutzen-2006), along with a number of more recent publications: Bengtsson-2006; Kiehl-2006; Cicerone-2006; MacCracken-2006; Lawrence-2006; Carlin-2007a; Carlin-2007b; Schelling-2007; Cascio-2008; Barrett-2008; Robock-2008a; Robock-2008b.

[3] Cicerone-2006, page 227.

[4] Ibid.

[5] Barrett-2008.

By clearly addressing technical questions regarding what research would be necessary to maximally reduce uncertainties, the study's aim is to better inform the ongoing debate about whether (or what types of) climate engineering research paths *should* be pursued.

Point 2 above (that climate engineering can be perceived as a substitute for mitigation) is so central to any discussion about climate engineering that it demands further elaboration. In this report, the focus is on climate engineering concepts that could serve as a rapid palliative response to near-term *climate emergencies* (defined in Section 1.1); this study explicitly does not consider the use of climate engineering as a long-term alternative to GHG emission reductions. This focus enables technical rigor and avoids discussion of socio-political issues beyond our expertise.

Based on the technical evaluations presented, the authors of this report believe that the extent to which shortwave climate engineering concepts would actually alter the perceived need for emission reductions would depend largely upon the results of a research program such as that outlined in this report. If it became clear that the side effects of deployment would be severe or are inherently highly unpredictable, then these concepts would likely have very minimal bearing on emissions reduction efforts. However, if the research suggested that the immediate side effects of intervention would be small, additional technical issues discussed in this study—such as the incomplete offsetting of GHG-induced climate changes, ocean acidification (*discussed in Section 2.1*), and the drawbacks of long-term dependence on climate engineering (*discussed in Section 2.5*)—would still have an important bearing on the connection. The group therefore believes that the relationship between climate engineering and $CO_2$ policy is too complex and multi-faceted to currently assign clear directionality between the encouragement of climate engineering research and the discouragement of $CO_2$ mitigation policies.

## *Preface*

This report is the result of a one-week study sponsored by the Novim Group, a non-profit corporation dedicated to analyzing global problems from a singularly scientific perspective.

Novim's mission, simply stated is:

- *To provide clear scientific options to the most urgent problems facing society*
- *To explore and explain the feasibility, probable costs and possible consequences of each course of action*
- *To distribute the results without advocacy or agenda both quickly and widely*

The currently emerging scientific discourse and growing public awareness surrounding the broad subject of climate engineering made it an ideal subject for the Group's first study. Novim's founders and this study's participants feel that public and professional discourse would benefit from a broad technical analysis of the subject.

The study methodology was one that has proven productive in the past. A group of both subject-matter experts and technically informed non-experts was assembled for intense discussion, debate, and writing of draft material. The draft materials were then edited and expanded into this report, which has been approved by all the participants and is published by the Novim Group. Edited portions and summaries of this report have been presented at scientific conferences, and may be submitted to one or more scientific journals. The content will also be condensed and clarified for a broad non-technical public audience, and presented through a wide range of popular media formats.

## *A Note on Conflicts of Interest*

Participants in the Study attended this workshop as individuals, based on a shared strong desire to wrestle with new and daunting questions that all expect will grow in importance to society. The ideas and opinions expressed herein do not necessarily represent those of any institutions or organizations with which any of the participants are individually affiliated.

It is the group's assessment that no individual brings a conflict of interest, either personal or professional, to this work. However, all felt it important to directly address the possibility that readers might perceive a conflict of interest on the part of the convener, Dr. Steven E. Koonin. During the meeting and preparation of this report, Dr. Koonin was the Chief Scientist at BP plc, one of the world's largest international oil companies. At first glance, some readers may perceive anyone working at any oil company to have an interest in distracting society from the job of reducing global $CO_2$ emissions, since the use of their products creates these emissions. Correspondingly, they may expect any investigation of climate engineering to be positively biased, aimed at distracting the World from getting on with the job of managing carbon emissions.

The study group acknowledges the plausibility of this line of reasoning. However, in this instance, Dr. Koonin has an extensive history of devoting a part of each year to small-group studies of societally-relevant science. These activities long predate Dr Koonin's joining BP, and BP allowed him to continue this practice in his individual capacity. BP contributed no funds into this study and had no influence over its content. Moreover, as discussed in the report Prelude, all participants share the belief that the relationship between climate engineering and $CO_2$ policy is so complex and multi-faceted that directionality cannot straightforwardly be assigned between encouragement of climate engineering research and discouragement of $CO_2$ reduction policies.

# 1 Introduction and Context

## 1.1 The Possibility of Climate Emergencies

The recent Fourth Assessment Report (AR4) of the *Intergovernmental Panel on Climate Change* (IPCC) concluded with "*very high confidence*" that anthropogenic accumulations of greenhouse gases (GHGs) in the atmosphere are affecting the Earth climate.[6] AR4 also documented a range of scientifically measured impacts on human and natural environments due to the current warming. As anthropogenic GHG concentrations continue to rise, their environmental impacts will increase.

However the sensitivity of the climate system to continued increases in atmospheric GHG concentrations remains significantly uncertain. AR4 placed the "*likely*" (66% likelihood) global mean temperature increase due to a doubling of $CO_2$ from pre-industrial levels at between 2ºC and 4.5ºC, while also noting that "values substantially higher than 4.5°C cannot be excluded."[7] This very broad range means that it is not currently possible to define a "safe" level of atmospheric $CO_2$,[8] because the future rate, magnitude, and distribution of climate change impacts are all far from certain. There is no guarantee that the previously observed trend of gradually intensifying consequences will continue. Indeed, the significant reduction in arctic summer sea ice coverage during the summers of 2007 and 2008 (~30% lower coverage than predicted by recent linear trends)[9] highlights the possibility that non-linear feedbacks in the climate system could accelerate these impacts.[10,11]

It is possible that international efforts to stabilize $CO_2$ concentrations will be sufficient to prevent or delay the worst climate impacts, that the world will warm only another 2°C or 3°C, that ice sheets will melt slowly, and that most of the consequences will be gradual, allowing smooth adaptation by human societies and natural ecosystems. However, it is also possible that even with concerted near-term efforts and cooperation on emissions reductions, the climate system could change quickly and unexpectedly, or could change smoothly, yet with consequences that are both severe and unanticipated. For example, the Greenland and West Antarctic ice sheets could disappear more rapidly than currently projected and the impacts of a warmer world on humans and natural ecosystems—ranging from rapidly increasing sea level to more frequent and stronger storms—could be more severe than current median predictions.[12]

We define *climate emergencies* as those circumstances where severe consequences of climate change occur too rapidly to be significantly averted by even immediate mitigation efforts. Given the long lifetime of $CO_2$ in the atmosphere, even aggressive near-term mitigation efforts would take several decades to stabilize and begin reducing atmospheric $CO_2$ concentrations.

---

6 IPCC-2007b

7 Ibid, page 38.

8 Apart from the impossible (in the near-term) goal of the pre-industrial level of 280ppm.

9 NSIDC-2007.

10 For one example of a postulated positive climate feedback related to the loss of arctic sea ice, see: Lawrence-2008.

11 The IPPC AR4 refers specifically to the possibility of "large-scale singularities" in the climate system that rapidly increase the rate and impacts of climate change: IPCC-2007b, page 65.

12 It is beyond the scope of this report to discuss the myriad possibilities for such climate change impacts. For a comprehensive review of potential impacts see: IPCC-2007a.

Consequently, mitigation can at best ameliorate consequences of climate change only over a similar multi-decade period.

While offering no predictions about the likelihood of either climate emergencies or gradual climate change (or anything in between),[13] we recognize that there is now a non-negligible possibility of a climate emergency in one form or other. Serious consideration of potential responses to such an eventuality is therefore important and reasonable.

### *1.2 Climate Engineering as a Potential—but Uncertain—Response to Climate Emergencies*

Intentional intervention in the climate system has long been discussed as one approach to deal with the effects of rising GHG concentrations. Such "climate engineering" concepts (also known as "geoengineering") vary widely in design and specific objective. Existing concepts range from those aimed at directly reducing atmospheric GHG concentrations, to those targeted at averting the consequent global warming by reducing the amount of solar radiation absorbed by the Earth, and even include some aimed at simply altering the regional to global scale distribution of climate change impacts.[14,15]

This study considers only climate engineering concepts that might serve as rapid palliative responses to global climate emergencies (*i.e.* having a significant palliative impact within several years of deployment.) This eliminates concepts aimed at directly reducing atmospheric GHG concentrations (*e.g.* large-scale reforestation, ocean-iron fertilization, atmospheric scrubbing of $CO_2$, *etc.*) because, similar to mitigation efforts, they would take several decades to have a substantial impact on the climate system.[16]

In contrast, concepts for modestly reducing the incoming shortwave solar radiation absorbed by the Earth would have a prompt cooling impact on the climate system.[17] Moreover, previous investigations suggest that the direct economic costs to deploy any one of several such *shortwave climate engineering* (SWCE) concepts could be very low by international standards (order of $30B annually, or ~0.05% of current global GDP.)[18]

We currently understand very little about either the potential utility or the risks of reducing absorbed solar radiation. This limited understanding stems most immediately from the very limited number of technical investigations that have been conducted to date, but also, and perhaps irreducibly, from the inherent complexity of the climate system and biosphere. Basic understanding is sufficient to conclude that a simple decrease of the solar constant can compensate for the increase in global average temperature caused by anthropogenic GHG emissions, but would not fully eliminate all impacts of climate change. For instance, temperature compensation at the regional scale would not be perfect, and other climate parameters currently perturbed by atmospheric GHGs—most notably ocean acidity—would remain largely unaffected. But beyond these basic observations, we know essentially nothing about the net combined impacts of shortwave climate engineering and elevated GHGs on a wide range of other climate and ecological parameters (*e.g.* regional precipitation, atmospheric and oceanic circulation, patterns of interannual variability, net ecological productivity, etc) or about the extent to which various SWCE concepts might be optimized to address them.

[13] Readers are referred to IPCC AR4 for recent consensus estimates of scenario probabilities: IPCC-2007b.

[14] For a basic taxonomy of climate engineering (or geoengineering) concepts, please see Figure 1 on page 12.

[15] For a detailed review and history of climate engineering concepts up to 2000 see: Keith-2000.

[16] Significant technical difficulties are associated with rapidly removing large quantities $CO_2$ for the atmosphere. Proposed approaches for removing carbon dioxide from the atmosphere include concepts such as enhancing biological uptake through fertilization or chemical removal and sequestration with engineered systems. However, even if such efforts were to become economically competitive with low-carbon energy systems, the scale required to make a significant impact on the existing stock of carbon in the surface ocean-atmosphere system is so large that reducing the atmospheric carbon dioxide concentration by a significant amount more quickly than several decades is essentially impossible.

[17] Natural events such as the Mt. Pinatubo eruption in 1991 have conclusively demonstrated the immediacy of such cooling impacts due to a temporary increase in the Earth's albedo. See, for example: Soden-2002.

[18] NAS-1992; Schelling-1996; Keith-2001; Barrett-2008.

This large uncertainty makes the near-term deployment of any full-scale SWCE concept highly risky—unanticipated negative impacts on human and ecological systems could overshadow the expected benefits. Indeed it is this high level of risk that has largely deterred serious consideration of these concepts to date (as discussed above in the Prelude to this report.)

But given the possibility of extreme climate emergencies discussed above, it is important that the best technical information and judgment be applied in assessing the risk-effectiveness of deploying any SWCE concept. Our study is a first attempt to outline a research agenda that would clarify and might reduce the uncertainties in such an assessment.

### *1.3 Reducing the Uncertainties: The Focus and Limited Scope of this Study*

Our weeklong study focused on the question:

> **"What program of scientific and engineering research over the next decade would maximally reduce the technical uncertainties associated with potential shortwave climate engineering responses to climate emergencies?"**

Balancing limited study time against a desire for depth and rigor, we focused our detailed technical analyses exclusively on the injection of aerosols[19] into the stratosphere to increase the Earth's shortwave albedo (reflectivity.) Previous evaluations of such concepts—particularly those involving the stratospheric injection of sulfate aerosols similar to those from volcanic eruptions—have already suggested the basic engineering and economic feasibility of such ideas.[20] Moreover, a recent body of preliminary scientific investigations[21] of this family of concepts provides a solid foundation upon which to define a comprehensive research program for reducing uncertainties.

Our focus on stratospheric aerosols does not imply that we consider it to be the only viable shortwave climate engineering concept worthy of consideration (although, with our present state of knowledge, we do judge it among the most feasible and potentially useful.) Rather, our detailed evaluation of this one family of concepts aims to provide both: (1) general insights into some issues related to *any* shortwave climate engineering proposal (presented mainly in Part 2 of this report); and (2) an example and template of the kind of systematic program necessary to reduce uncertainties in other types of potential interventions (presented mainly in Part 3 of this report.) Examples of alternative SWCE concepts are included with references in Figure 1.

---

[19] We consider in this study the full range of proposed (and possible) aerosols that could be injected into the stratosphere to increase the Earth's albedo.

[20] NAS-1992; Crutzen-2006.

[21] For references and a summary discussion of the recent scientific investigations, please see: Rasch-2008b.

# 2 Technical Considerations for Climate Engineering Interventions

## *2.1 The Potential Utility and Limitations of Shortwave Climate Engineering Interventions*

Basic understanding of the climate system's energetics suggests that reducing the shortwave (SW) solar radiation absorbed by the Earth could compensate for the global-scale warming induced by increasing GHG concentrations. Moreover, basic technical analyses suggest that the scale of the required reduction would be within our current technological capabilities. Beyond this simple thinking, however, several important considerations loom. In particular, humans and ecosystems exist within local environments, whose response to global radiative forcing can be notably different (and more complex) than the global average response. As a result, it is essential to consider the combined regional (and, if possible, even local) impacts of shortwave climate engineering (SWCE) and GHGs. Adding to this, most discussions of SWCE to date assume a globally uniform and constant reduction of the incident SW radiation through a modest (1-2%) reduction in the solar constant. Such "uniform SWCE" is convenient both computationally and conceptually, but may not be realizable, or even desirable, in practice. Spatial and temporal modulation of the injection processes, as well as engineering of aerosol lifetime (and perhaps transport), offer additional control parameters that might be useful for optimizing an SWCE intervention.

### *2.1.1 Large Uncertainties and Imperfect Offsetting of GHG-Induced Climate Change*

Current understanding of the climate system is already sufficient to conclude that shortwave climate engineering (SWCE) would not be a panacea for all GHG-induced climate change. For example, ocean acidification has been identified as a major consequence of increasing atmospheric $CO_2$ concentrations—the decreasing pH of the oceans could have severe consequences for ocean ecosystems and even impact other climate system feedbacks.[22]

As SWCE would not significantly influence $CO_2$ concentrations,[23] ocean acidification and its impacts will continue to increase with growing atmospheric $CO_2$ concentration, independent of any SWCE intervention.[24]

For climate emergencies however, complete cancellation of all GHG-induced climate change is unlikely to be the primary objective. Rather, amelioration of the most immediate and severe consequences of GHG-induced climate change will likely be the dominant concern, and a reasonable body of evidence indicates that SWCE has the potential to diminish some severe climate impacts of increased GHG concentrations—particularly those associated with increasing average global temperature. This evidence includes aspects of basic theory (*see Box 2.1.2.1*), preliminary climate model simulations (*see Box 2.1.2.2*) and observations of related natural experiments (*see Box 2.1.2.3*.) In particular, these considerations collectively provide high confidence that modest reductions of the shortwave solar radiation absorbed by the Earth's atmosphere would be effective in cooling the Earth very rapidly (less than one year.)

---

[22] Royal-2005.

[23] A second-order effect of SWCE on $CO_2$ concentrations may exist due to changes in the net global ecological productivity, however this influence is negligibly small relative to current anthropogenic contributions to $CO_2$ concentrations.

[24] Other climate engineering (*a.k.a.* geoengineering) concepts, unrelated to the shortwave concepts examined herein, have been proposed for counteracting ocean acidification. The 2005 Royal Society review of ocean acidification briefly examined these proposals and discussed their potential and limitations. Royal-2005, page 37.

Beyond this basic understanding of the effects on global temperature, very little is understood regarding the temporal and spatial impacts of SWCE on a broad range of other climate parameters. Some recent climate simulations (*see Box 2.1.2.2 and Figure 2*) suggest that uniform SWCE might reasonably offset the spatial distribution of GHG-induced temperature changes associated with a doubling of $CO_2$, while other simulations suggest a limited potential of uniform SWCE to simultaneously compensate both temperature and precipitation changes (*discussed in Box 2.1.2.2*.) However these studies have used idealized climate models, leaving out exploration of possible SWCE impacts on important climate parameters and feedbacks. For example, these studies did not use models that capture natural patterns of climate variability (such as ENSO and NAO.) As discussed in Box 2.1.2.4, the sensitivity of these natural patterns of climate variability to small changes in the climate system suggests that these patterns might be significantly altered by SWCE. A range of other important climate parameters and feedbacks (such as net ecological productivity and its feedback on the global carbon cycle) also remain largely unstudied to date. The limits of these studies prevent firm conclusions being drawn for existing data regarding the spatial and temporal impacts of SWCE on even temperature and precipitation patterns.

Even in the equilibrium simulations that show good compensation of GHG-warming by uniform SWCE, careful examination of Figure 2 shows that the spatial distribution of temperature and precipitation compensations is imperfect at doubled $CO_2$ levels and that these imperfections grow with greater GHG-loadings (*i.e.* a quadrupled $CO_2$ relative to pre-industrial levels.)[25] Indeed, independently of whether these simulations accurately capture the spatial (regional) compensation ability, the compensation ability of uniform SWCE will decrease with increasing GHG concentrations. It is also possible that SWCE could *amplify* other regional consequences of GHG-induced climate change. For example, the drying of the Sahel region observed during the second-half of the 20$^{th}$ century has been tentatively attributed to the combined, reinforcing impacts of aerosol-induced cooling and GHG-warming (*see Box 2.1.2.5*.)

A final issue, raised by the natural experiment of the Mt Pinatubo eruption, is that of transient versus equilibrium climate responses to SWCE. As discussed below (*see Box 2.1.2.3*), if the stratospheric aerosol loading created by that eruption had been sustained indefinitely, the long-term equilibrium cooling of climate could have been up to six times the 0.5ºC cooling observed in 1992. But the short lifetime (one to two years) of the sulfate aerosols in the atmosphere, combined with the climate's thermal inertia (caused predominantly by the large heat capacity of the oceans), reduced their cooling impact.

SCWE interventions would likely be of a different character—a smaller stratospheric aerosol loading sustained for a much longer duration. There would also be longer time scales associated with ramp-up during deployment, as well as possible temporal modulation to any "steady state" SWCE intervention. As a result, it is important to understand the temporal response of the climate system, and what natural phenomena like the Pinatubo eruption can (and cannot) teach us. The limited investigations of SWCE to date have focused predominantly on long-term equilibrium impacts, leaving considerable uncertainty about shorter-term temporal behavior.

[25] Govindasamy-2003.

This brief review indicates that SWCE could not perfectly reverse effects of GHGs, but that such climate engineering has the potential to diminish some of the severe consequences associated with GHG-induced warming. However because the couplings between SWCE and a range of climate (*e.g.* patterns of interannual variability, atmospheric chemistry) and ecological parameters (*e.g.* productivity and species fitness) are largely unstudied, there remain major uncertainties regarding the combined net climate impacts of SWCE and elevated GHG concentrations.

### *2.1.2 Useful for Some, but Not All "Climate Emergencies"*

It is beyond the scope of this report to review the myriad scenarios involving rapid or accelerating climate change consequences that could constitute a "climate emergency"—a range of such scenarios are reviewed in other assessments[26] and discussed in the literature.[27] However, it is important to note that different climate emergencies would call for different responses—and that SWCE might not be an effective response for many cases. For scenarios where SWCE could be an effective response, the nature of the emergency would define the timescales over which the intervention would need to have an impact and the magnitudes of the targeted corrective climate response, and will thus strongly influence what might be considered the "optimal" attributes of an SWCE intervention and control system.

One type of climate emergency for which a SWCE response might be considered are those generated by an increased expectation of future severe climate consequences that even ambitious mitigation efforts appear insufficient to avoid. Examples of such future consequences might include an increase in the decay rate of the Greenland ice-sheet such that projected sea-level rise would seriously threaten coastal communities in the next several decades, or a rapid increase in the rate of permafrost thawing and GHG release resulting in an acceleration in the rise of atmospheric GHG concentrations. In these scenarios, severe climate change consequences would become strongly anticipated and appear unavoidable without a near-term climate engineering intervention, but still remain in the future; hence, they would likely call for slower development and deployment timescales, in order to allow time for a wider diversity of tests that could help determine reliability, side effects, and even comparison among various technology options.

Another class of climate emergencies consists of those where severe consequences begin manifesting immediately. In these scenarios, the hope would be that a climate engineering intervention would reverse (or at least diminish or slow down) these consequences. Examples of these might include increases in extreme weather (manifest as persistent occurrence of large storms or heat waves) with significant impact on society and infrastructure, as well as severe and prolonged droughts. In these scenarios, because significant harm to society is already taking place, more immediate responses will likely be called for.

Different climate emergencies would also call for varying magnitudes of response. For example, concern about trends in the melt rates of mountain snow and ice cover, along with the resulting hydrological impacts, might call for an effort to cancel the warming effects of any further increases in GHG concentrations. On the other hand, a dramatic acceleration of the melting of the Greenland Ice Sheet might call for a much larger response aimed at lowering Arctic temperatures sufficiently to restore the summer Arctic sea ice—which might even require lowering the global average temperature significantly relative to what societies have experienced in the latter 20th century, and possibly even below pre-industrial levels.

---

[26] See, for example: IPCC-2007a.

[27] Lenton-2008.

For scenarios where the dominant climate impacts are regional rather than global (as is the case for many of the above examples), there is also the possibility of regionally targeted SWCE. However, similar to the way in which the differential spatial distributions of aerosol and GHG forcings likely interacted to generate the Sahel drought (*see Box 2.1.2.5*), SWCE to stabilize climate in one region might amplify the impact of increasing GHG concentrations elsewhere. A similar challenge exists in attempting to stabilize a subset of global climate parameters, as other climate parameters will be simultaneously impacted. For example, the results of Bala *et al* (2008)[28] (*discussed further in Box 2.1.2.2*) suggest that the simultaneous exact cancellation of GHG-induced temperature and precipitation changes is not possible.

Finally, some climate emergency scenarios may not be at all responsive (or at best minimally responsive) to an SWCE intervention. SWCE could directly and immediately reduce the shortwave solar radiation reaching the Earth's surface, inducing rapid (<1yr) global cooling. However, if the Ross Ice Shelf were to break off and the West Antarctic Ice Sheet were to begin sliding into the Southern Ocean, it is not clear whether—once started—such complex glacial dynamics could be halted (or even significantly delayed) by even an immediate reduction in global average temperatures. A variety of similar tipping points have been discussed in the literature,[29] and once a threshold is crossed in the climate state, the impacts could be irreversible. At that point, reducing the surface temperature, even abruptly, might have no effect on the consequences. This consideration suggests the possibility that "pre-emptive" SWCE to avoid crossing such a threshold might be risk-effective. However, far better understanding of both the risks associated with SWCE and the expected likelihood and consequences of crossing such a threshold will be required before such a decision could be taken confidently.

Present uncertainties about the temporal, spatial, and climate parameter influences of SWCE interventions prevent confident assertions about the range of climate emergencies for which SWCE could be a useful response. However the simple examples considered in this section emphasize the importance of these issues for the design of any SWCE system. Moreover, our discussion further emphasizes the value of extensive research covering the full range of possible climatic responses to SWCE, so as to best understand the utility and limits of any potential SWCE system before it might be called upon to address a climate emergency.

[28] Bala-2008.

[29] See, for example: Lenton-2008.

**Box 2.1.2.1 Basic Concept of Shortwave Climate Engineering**

The Earth's average global temperature will be steady if the solar energy absorbed by the Earth is balanced by the thermal radiation energy leaving the Earth.[30]

As the Earth is warmed principally by absorbing shortwave (<3 μm) solar radiation and cooled by radiating longwave infrared (~8-14 μm) radiation to space, variations in the balance of incoming shortwave and outgoing longwave radiation can heat or cool the Earth. For example, GHGs absorb and re-emit some of the outgoing longwave radiation, preventing this radiation (heat) from escaping the Earth's atmosphere. Absent any change in the absorbed shortwave radiation, the direct effect of GHGs is to produce a downward energy flux at the top of the atmosphere, leading to additional heating of the Earth's surface.

The basic concept of SWCE is to reduce the shortwave radiation absorbed by reflecting more of it back to space (~30% is already reflected by the natural constituents of Earth's atmosphere and surface.) Basic considerations show that an additional 1-2% of reflectivity would balance, in net energy terms, the additional heating caused by a doubling of atmospheric $CO_2$ concentration. The globally average longwave radiative forcing due to a doubling of atmospheric $CO_2$ is approximately 4 $W/m^2$. As the globally averaged shortwave solar energy reaching the Earth is roughly 342 $W/m^2$, the reflection of an additional ~1% by SWCE would roughly restore the energy balance. Figure 1 provides a simple schematic overview of the Earth's radiation balance, and indicates where various SWCE schemes would alter the flow of shortwave solar radiation.

However, these simple considerations ignore the distribution of energy within the climate system. As the incoming shortwave and outgoing longwave energies have different spatial and temporal distributions, the net impact of GHG+SWCE on the climate system would not be zero. Climate features such as regional temperatures and precipitation levels, interannual variability, ecological productivity, and many others could all remain impacted. Scientific investigations of these distributional issues using observation of natural experiments (*see Box 2.1.2.3*), climate modeling (*see Box 2.1.2.2*), and potentially even field testing of SWCE (*see Section 3.2*) are needed to provide insight.

[30] For simplicity, we set aside the minor fluxes associated with heat flowing from the solid Earth, and ignore the small net amount of additional energy stored each year as chemical, potential, or kinetic energy. The net energy stored by these processes is small by comparison to the heat store in the oceans or the cosmic 3K background radiation.

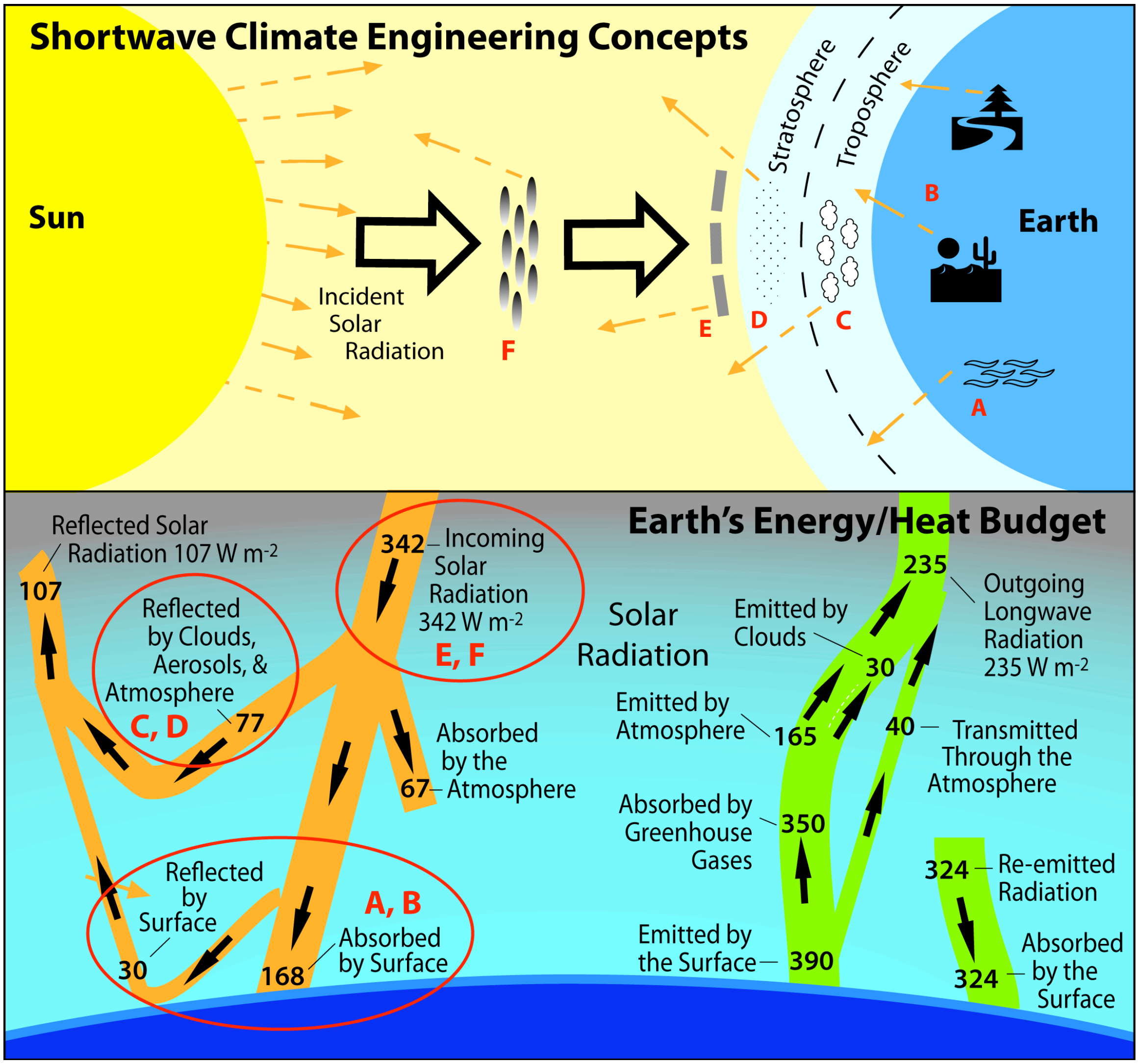

***Figure 1*** Schematic overview of Shortwave Climate Engineering (SWCE) concepts

*Top:* Schematic depiction of the physical location of a range of proposed SWCE concepts. Discussion and references for most concepts can be found in Keith (2000)[31]—select recent references have been noted where applicable.

- **A** *Increasing ocean surface reflectivity (e.g. microbubbles)*
- **B** *Increasing land surface reflectivity (e.g. crop modification)*
- **C** *Cloud whitening*[32]
- **D** *Stratospheric aerosols* (discussed in detail, with references, in Part 3 of this study)
- **E** *Orbital mirrors or reflectors*
- **F** *Reflector arrays at the Lagrange (L1) point*[33]

*Bottom:* Schematic of the Earth energy budget, depicting the fluxes associated with both shortwave solar radiation (left) and Earth's longwave thermal radiation (right.) The left portion of the diagram also associates SWCE concepts with the component of the shortwave energy flux that they would directly impact. Derived from figure in Kiehl and Trenberth (1997.)[34]

---

31 Keith-2000.

32 Latham-2008; Salter-2008.

33 Angel-2006.

34 Kiehl-1997.

**Box 2.1.2.2 Initial Climate Model Simulations of Shortwave Climate Engineering**

The first published climate model simulations of SWCE were performed by Govindasamy and Caldeira (2000),[35] and have been recently expanded by Caldeira and Wood (2008.)[36] These simulations of uniform SWCE used a decreased solar constant to crudely represent the radiative effects of SWCE, and a doubled carbon dioxide concentration relative to pre-industrial levels. Both studies found that a uniform reduction in solar intensity significantly reduced both temperature and precipitation changes from increased $CO_2$ that were statistically detectable[37]—for example, see Figure 2. This near-cancellation of the effects of GHGs occurred both seasonally and regionally.

Caldeira and Wood (2008)[38] conjectured that feedbacks in the climate system could help explain this cancellation. Because the poles experience a lower average insolation and have a high albedo due to the ice cover, a reduction in the solar constant has a smaller direct effect there than at lower latitudes. However, snow and ice albedo feedbacks operate more strongly near the poles, making these regions sensitive to small variations of incident radiation. Moreover, while the radiative forcing from a doubling of atmospheric $CO_2$ is somewhat less than 4 $W/m^2$, the locally absorbed solar radiation can change by more than 100 $W/m^2$ depending upon whether there is surface water or sea ice. Additionally, as sea ice insulates the atmosphere from the underlying ocean, the thermal energy transfer between the ocean and atmosphere can also change by more than 100 $W/m^2$. Thus, the character of regional climate feedback responses can dominate the contributions from changes in radiative forcing alone, and these regional effects can even dominate the global climate response. As a result, the net behavior of the climate system could, to first order, be nearly independent of the specific details of the climate forcing. In the Caldeira *et al* simulations, reducing the incident solar radiation allows the sea ice in a high-GHG world to return nearly to its pre-GHG coverage. When this occurs, high latitude seasonality also returns close to its normal pattern.

However other recent simulations by Bala *et al* (2008)[39] have suggested limits to the compensation potential of uniform SWCE. Their results show that the global hydrological sensitivities[40] to elevated GHGs are significantly different from those to solar intensity, and attribute this to the greater sensitivity of surface evaporation rates to shortwave versus longwave radiation fluxes. These results suggest that GHG temperature and precipitation changes could not simultaneously be precisely compensated by solar intensity reductions. For example, eliminating temperature changes would be expected to produce a net reduction in global precipitation. The simulations of Caldeira *et al* (discussed above) are qualitatively consistent with these results—examination of Figure 2 shows that temperature is (on average) slightly under compensated and precipitation is (on average) slightly over compensated, indicating greater shortwave than longwave hydrological sensitivities. However quantitative differences between the two simulations are illustrative of the discrepancies (or uncertainty) in the estimates of how well solar intensity reductions could compensate for GHGs.

---

[35] Govindasamy-2000.

[36] Caldeira-2008.

[37] With 95% confidence based on a 30-year climatology.

[38] Ibid.

[39] Bala-2008.

[40] The hydrological sensitivity is "defined as the percentage change in global mean precipitation per degree warming." Ibid, page 7664.

Both the Caldeira *et al* and Bala *et al* studies are idealized models of an SWCE intervention, in that the solar constant is simply reduced. The precise spectral, spatial and temporal radiative impacts of "real" SWCE concepts were, therefore, not captured by these simulations. Recent modeling efforts by Rasch *et al* (2008)[41] and Robock *et al* (2008)[42] have begun to address this issue by directly simulating the addition of sulfate aerosols to the stratosphere. While some important parameters (such as aerosol particle size distributions) are artificially fixed in the simulations,[43] both studies use a model of the atmospheric sulfur cycle to calculate stratospheric aerosol concentrations as a function of injection rate and time, and then use those concentrations to determine radiative impacts. The results of these studies show some differences from the uniform solar reduction studies (particularly in details of the spatial distribution of aerosol cooling),[44] however the general conclusions are similar—reductions in radiative forcing caused by the sulfate aerosols appear, to a large extent, to cancel the temperature and precipitation changes caused by increasing GHGs.

All of these first-generation modeling studies of SWCE should be viewed as exploratory. Beyond the small number of climate models used (and the simple climate process models used in some studies—such as a slab ocean or static sea ice), these studies have also focused primarily on the long-term equilibrium response of the climate system to an SWCE intervention. As illustrated below (*see Box 2.1.2.3*), the transient climate response to a reduction in solar radiation can be significantly different. Moreover, none of these studies have examined the potential interaction between SWCE and the natural patterns of climate variability, such as the North Atlantic Oscillation (NAO) and the El Nino/Southern Oscillation (ENSO); see Box 2.1.2.4 for further discussion.

[41] Rasch-2008a.

[42] Robock-2008c.

[43] See Box 2.3.1.2 for discussion of uncertainty regarding sulfate aerosol particle sizes.

[44] Robock-2008c emphasizes several regional examples where cancellation is imperfect.

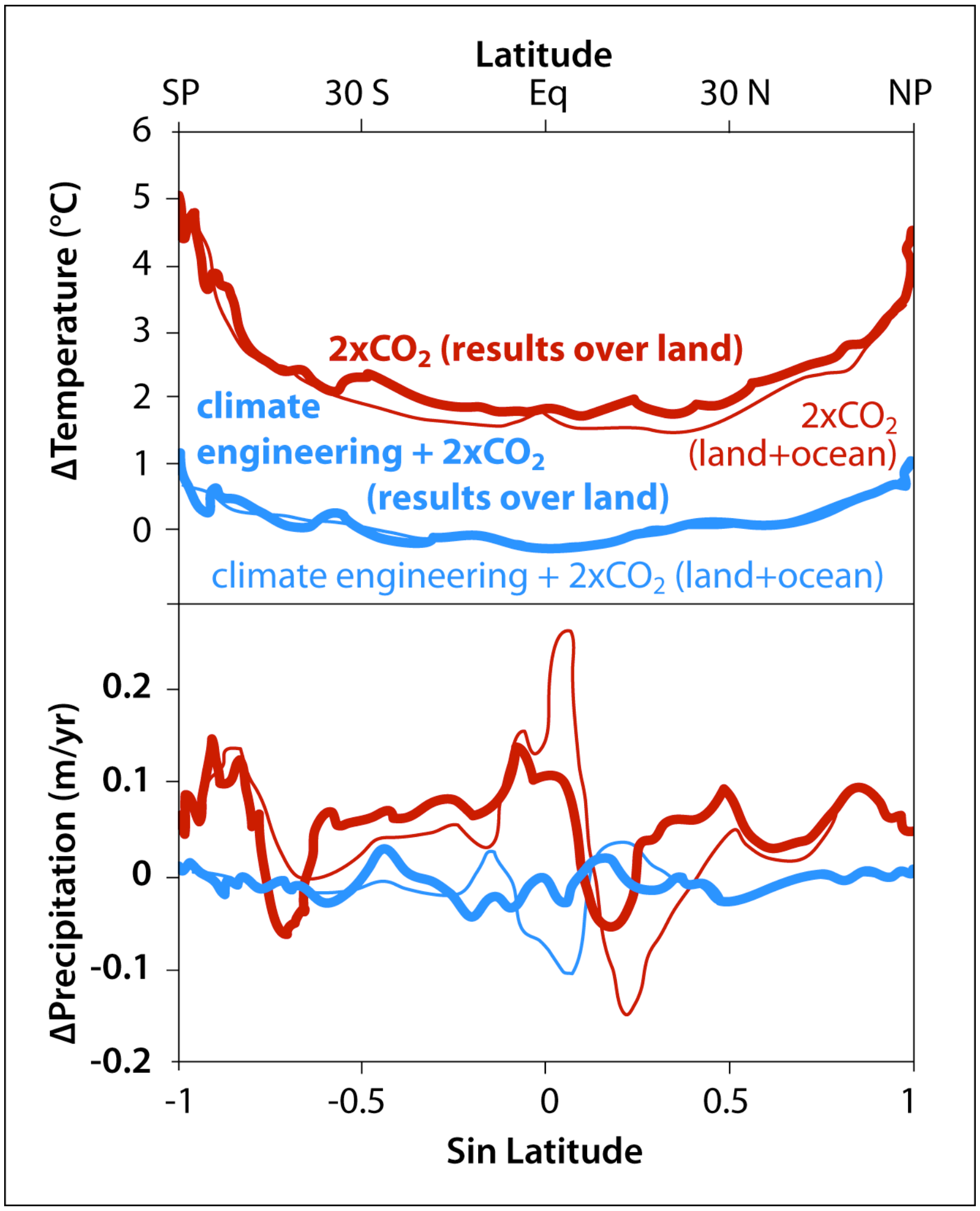

## *Figure 2* Initial climate model results for a globally uniform SWCE scenario

Climate model results averaged by latitude for a highly idealized SWCE scenario described in detail in Caldeira and Wood (2008.)[45]

Changes in temperature are shown in the top panel. Changes in precipitation are shown in the bottom panel. Results averaged over land only are shown in thick lines and results averaged over the entire latitude are shown in thin lines.

Red lines (*$2{\times}CO_2$*) show climate model results for a doubled atmospheric $CO_2$ scenario relative to a control simulation.

Blue lines (*climate engineering*) show results for doubled atmospheric $CO_2$ content with a uniform 1.84% reduction in the sunlight incident on the Earth relative to a control simulation. In the control simulation with pre-industrial $CO_2$ concentrations, precipitation averages 1.02 m/yr globally and 0.96 m/yr over land.

45 Caldeira-2008.

**Box 2.1.2.3 "Natural Experiment" Parallels of Shortwave Climate Engineering**

Several historic volcanic eruptions—Tambora in 1815 (preceding the "Year Without a Summer" in Northern Europe and the Northeastern US in 1816), Krakatau in 1883, El Chicón in 1982, and Pinatubo in 1991—have been associated with short-term (~1 to 3 year) subsequent hemispheric cooling. It has been generally accepted that the eruptions caused the cooling (and spectacular sunrises and sunsets) by injecting aerosols into the troposphere and lower stratosphere, and that these effects diminished as the aerosols were removed from the atmosphere by sedimentation or scavenging by hydrometeors.[46]

As the most recent event, Mt. Pinatubo is also the most thoroughly studied of these major volcanic eruptions.

Figure ***3*** shows the correlation between increased atmospheric aerosol concentration from the eruption (measured by aerosol optical thickness) and the consequent average global cooling. In 1991, Pinatubo injected about 10 million tons of sulfur (10 TgS) into the stratosphere in the form of $SO_2$ (which is ~20% of annual global tropospheric sulfur emitted as $SO_2$ by the burning fossil fuels[47].) Ground based measurements have shown that at peak loading (immediately following the eruption) the shortwave planetary albedo very briefly increased by 0.02 above its "normal" value of about 0.30 as a result of the stratospheric aerosols formed, equating to a ~5% net decrease in the amount of sunlight reaching Earth's surface.[48] At the same time, the peak stratospheric loading also converted an additional ~20% of the total incident sunlight from direct to diffuse illumination (for a clear sky), causing an increase of up to ~300% in the amount of clear sky diffuse sunlight reaching Earth's surface at some locations.[49] As the aerosol loading of the stratosphere diminished to background levels over several years, the corresponding effects also disappeared with no evidence of a lasting impact.[50]

As a result of the Mt Pinatubo eruption, the 1992 global mean temperature of the Earth decreased by about 0.5˚C. This cooling has been simulated with a fairly high degree of fidelity in several different climate models, using estimates of sulfur injection or the radiative effects of that sulfur as the forcing variables. However, the thermal inertia due to the oceans' large heat capacity smoothes and delays the climate response to a change in radiative forcing. As a consequence, had the Pinatubo stratospheric aerosol loading been sustained indefinitely, the cooling would have been some six-times greater than the transient 0.5ºC observed.

This "natural experiment" provides strong evidence that stratospheric aerosols (and sulfate aerosols in particular) would diminish absorption of solar radiation by the Earth, and that the lifetime of the aerosols in the stratosphere is of the order of one to two years (although there can be substantial variation depending on aerosol particle size, latitude and altitude of injection[51].) Moreover, these natural experiments also illuminate the possible unexpected consequences of injecting aerosols into the stratosphere. For example, the eruption is thought to have diminished stratospheric ozone by about 3% on average (about 5% near the poles and 2% near the equator), while land plants are thought to have grown more vigorously after the eruption due to the increase in diffuse sunlight. (For example, Gu et al (2003)[52] have suggested that the increase in diffuse sunlight allowed more light to penetrate forest canopies, more than offsetting the effects of reduced direct and total sunlight.)

---

46 Robock-2000.

47 Stern-2005.

48 Ibid.

49 Ibid.

50 For examples of the scientific data and evaluations of the Mt. Pinatubo eruption see: Kinnison-1994; McCormick-1995; Soden-2002.

51 Rasch-2008b.

52 Gu-2003.

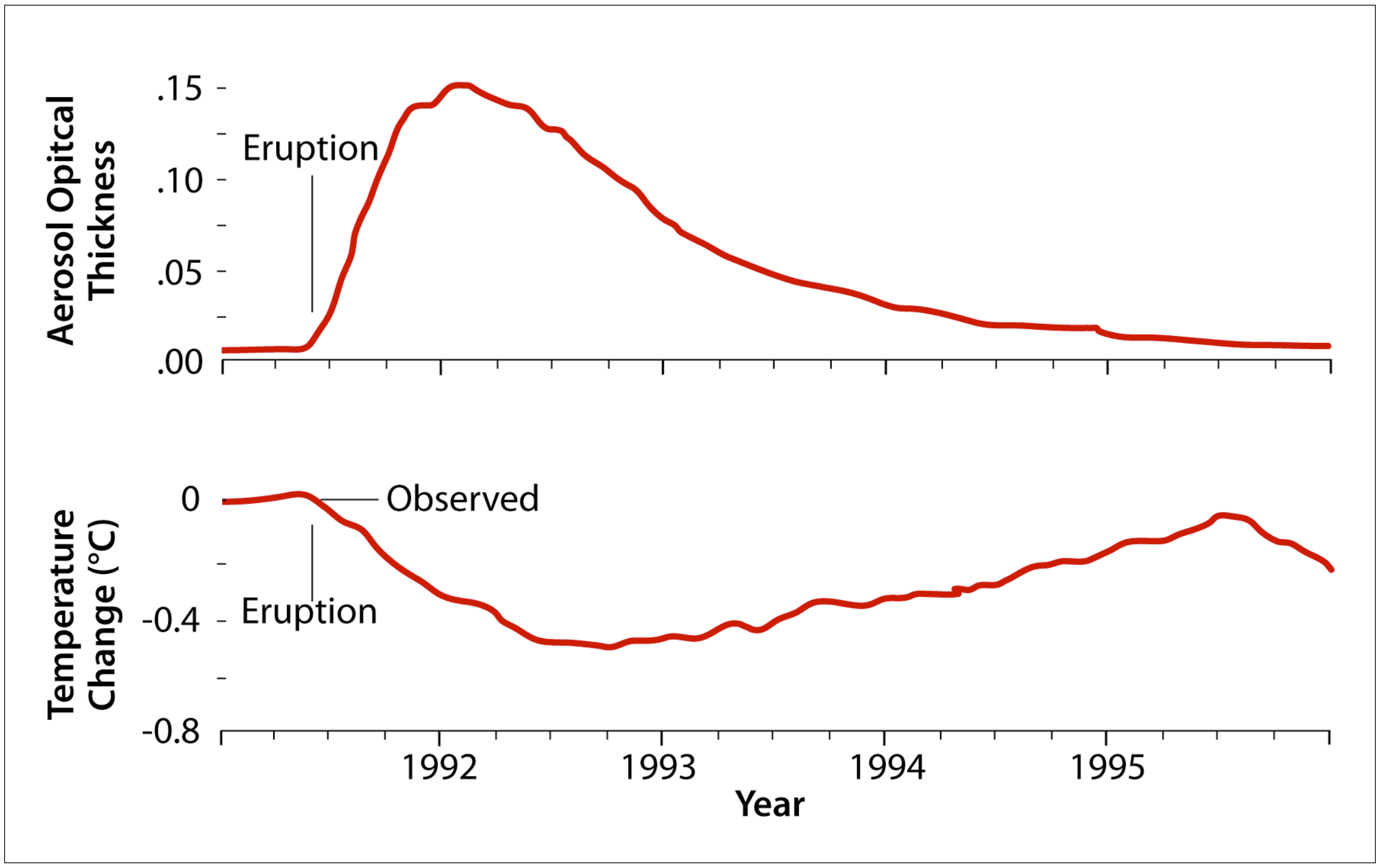

***Figure 3*** Global temperature impact of aerosols from the 1991 Mount Pinatubo eruption

Observed increase in atmospheric aerosols (top) and decrease in global average temperature (bottom) following the 1991 eruption of Mount Pinatubo. This image is derived from data and images available at the website of the NASA Earth Observatory.[53]

53 See for reference: http://earthobservatory.nasa.gov/Features/GlobalWarmingUpdate/

**Box 2.1.2.4 Sensitivity of Climate Variability to Shortwave Climate Engineering**

The amplitude and spatio-temporal structure of the natural patterns of climate variability are highly sensitive to changes in the climatological mean state of the atmosphere and ocean. Even small changes in the average radiative forcing of the global climate system—from either SWCE or GHGs (or the combination thereof)—could therefore result in large regional climatic changes. Well-known patterns of natural climate variability include the El Nino/Southern Oscillation (ENSO) and the North Atlantic Oscillation (NAO) phenomena, whose sensitivities both provide important examples for consideration.

In the case of ENSO, which accounts for most of the climate variability in the tropical Pacific and a substantial fraction in the global tropics and extra-tropical pan-Pacific, model results and theory[54] suggest that even small changes in the east-west difference in sea surface temperature or zonal wind stress in the tropical Pacific could produce large (up to ~100%) changes in the amplitude of ENSO, and cause zonal shifts in the location of the precipitation anomalies of up to ~3000km (which causes further changes in associated extra-tropical climate anomalies.) This high sensitivity suggests that ENSO might be strongly affected by SWCE. For example, if an SWCE intervention resulted in a net forcing of the tropical Pacific (*e.g.* 1 $W/m^2$), this might cause average changes that are sufficiently large (*e.g.* 0.5ºC in sea surface temperature) to have similarly large impacts on the amplitude and spatio-temporal structure of ENSO.

The NAO (along with several other leading patterns of climate variability in the mid-latitudes, such as the Southern Annular Mode) is the expression of the wobbling of the climatological jets due to interactions between storms and mean flow. The climatological position and strength of the jets—and hence the structure and amplitude of the NAO (and other patterns)—are sensitive to changes in the vertical and latitudinal structure of the diabatic forcing.[55] Hence, net changes in forcing due to SWCE might affect the climatological location of the jets and stormtracks, as well as year-to-year variability in the stormtracks.

The potential interactions between the natural patterns of climate variability and SWCE have not yet been explored. Given the potential sensitivity of these patterns, investigation of these interactions should be a central component of any SWCE research agenda (*see Section 3.1.2 for further discussion*.) These investigations should focus particularly on the combined impacts of increasing GHGs and SWCE, as patterns of climate variability could respond differently to the two climatic forcings.

[54] See, for example: Neelin-1998.

[55] See, for example: Hartmann-2000.

**Box 2.1.2.5 The Sahel Drought: Aerosol Cooling, GHG Warming, or Both?**

From the late-1950s to mid-1980s, the Sahel region of northern Africa experienced a lasting drought that had profound impacts on the human societies in that region. This regional drying has been attributed to changes of proximal tropical *sea surface temperatures* (SST), which altered regional evaporation and precipitation patterns.[56] As this period also coincided with both increasing atmospheric $CO_2$ concentrations and the largest historic tropospheric aerosol forcing due to human activities (fossil fuel and biomass burning), a number of climate simulations have looked explicitly at whether the SST changes were driven by GHG-induced warming and/or aerosol-induced cooling.

The consensus to date is that GHG-induced warming alone is insufficient to reproduce the historic regional drying. While some simulations with only GHG forcing did produce some regional drying in the Sahel, simulations that also included aerosol forcing replicated far better the historically observed drying.[57] The details of these simulations suggest that, while GHG-induced warming does have a notable impact on tropical SST, the differential spatial distribution of GHG-warming and aerosol cooling created spatial gradients in tropical SSTs that significantly amplified the Sahel drying.

The mid-20$^{th}$ century combination of high tropospheric aerosol loading and increasing atmospheric $CO_2$ concentrations bears some similarity to possible future SWCE+GHG "engineered" climates. This analogy—while highly imperfect—does highlight both the notion that uniform SWCE modeling can miss important impacts, as well as the possibility that an SWCE intervention (such as stratospheric aerosol loading) could *amplify* climate change impacts already being generated by increasing $CO_2$ concentrations.

It is also notable that, although a range of climate models with both GHG and aerosol forcing are able to reproduce the observed 20$^{th}$ century drying of the Sahel, these same models provide widely varying (sometimes opposite in sign) predictions for whether the 21$^{st}$ century trend of increasing $CO_2$ and decreasing tropospheric aerosol concentrations will lead to an increase or decrease of precipitation in the Sahel region.[58] This reality highlights the inherent complexity of the climate system, and the current challenges associated with accurate prediction of the impacts of any proposed climate engineering intervention.

---

[56] See, for example: Held-2000; Bader-2003; Biasutti-2006; Biasutti-2008.

[57] See, for example: Held-2000; Biasutti-2006.

[58] Biasutti-2008.

## *2.2 General Design Objectives for Shortwave Climate Engineering Systems*

The discussion above highlights three general design objectives for any SWCE system aiming to ameliorate the negative consequences of a climate emergency. Such a system should:

*(1)* Reliably induce targeted responses of the climate system;

*(2)* Have minimal anticipated undesirable impacts on untargeted components of the climate systems; and

*(3)* Have minimal potential for unanticipated consequences.

Each of these is an obvious and understandable goal, which gains greater specificity in the context of particular climate emergency scenarios. However the present uncertainty in SWCE climate impacts—coupled with natural complexity and variability of the climate system—creates the very substantial challenge of designing an SWCE system that addresses these goals.

Because any SWCE system is unlikely to fully achieve these three basic design objectives, we articulate two additional design objectives that provide some measure of risk mitigation:

*(4)* An SWCE system should be highly scalable to allow for sub-scale testing and a gradual deployment to full-scale.

Both sub-scale testing and gradual intervention ramp-up provide opportunities for careful monitoring to confirm predictions of targeted and secondary climate responses, and to detect and evaluate unanticipated consequences. (There are, however, notable limitations on the ability to detect the impacts in sub-scale testing, as discussed in detail in Section 3.2.)

*(5)* The expected climate impacts of an SWCE system should be reversible, ideally on a few-year timescale.

If predictability or unanticipated consequences prove to be problematic—even after several years of a full-scale intervention—the ability to rapidly "shut off" the climate impacts of an SWCE system is an important recourse. (Although the consequences of rapidly shutting off a long-standing SWCE intervention could be quite severe in their own right, as discussed further in Section 2.5.)

Given existing SWCE concepts and current uncertainties, a SWCE system design could not now be optimized to achieve all of these objectives simultaneously. Fully achieving even one objective appears unlikely (with the possible exception of 4), and judicious design tradeoffs will be part of any SWCE system development. A net risk assessment methodology, accounting for both the risks of SWCE and its palliative effectiveness, will therefore be essential.

Though cost-effectiveness could be considered as an additional design objective, we deem it of negligible importance relative to the others. Given the exceptionally low cost of developing and deploying a range of different SWCE systems,[59] the primary concern should be to achieve the objectives listed here, in order to lower the net expected risk of any intervention.

[59] NAS-1992; Schelling-1996; Barrett-2008.

### *2.2.1 A Note on "Natural Variability" as a Design Consideration*

The 1992 NAS review of climate engineering concepts (*a.k.a.* geoengineering) proposed that any intentional human intervention into the climate system should aim to stay "within the natural variability of the geophysical system."[60] This proposition rests upon the premise that, if climate engineering perturbations stay "within the bounds [previously observed] for natural experiments," such as volcanoes, they would likely "not produce instabilities or effects [in the climate system] that had not been produced before."[61] While seemingly sensible, there are several difficulties with the "natural variability" concept invoked by the NAS report.

Through its example treatment of the stratospheric aerosol-loading caused by volcanoes,[62] the NAS report implied that variations in the magnitude of specific climate parameters are a definitive indicator of natural variability. This magnitude-based interpretation of natural variability has since been used more explicitly by other assessments of stratospheric-aerosol injection concepts.[63] As pointed out in Box 2.1.2.3, however, the transient dynamics of the oceans' heat-capacity and the rapid decay of aerosol loading dominated the observed climate response to natural events. Consequently, the observed climate response to the large transient natural loading could differ significantly from that to smaller, but sustained, artificial loadings. A more robust definition of natural variability would include not only the magnitude of variation in specific climate parameters, but also the temporal and spatial scales of such variability. However, in such a multidimensional space, "staying within natural variability" looses analytic and operational clarity.

Finally, the NAS report itself pointed out that GHGs were already inducing changes that make historic natural variability "an incomplete guide to the future."[64] Indeed, for the climate emergency scenarios that we consider here, there is an assumption that one or several climate parameters would already be fundamentally different from the historical record. Even a natural event similar in magnitude and duration to previous natural events might not produce similar climatic response.

Examination of the historic "natural variability" of certain climate parameters does provide a good preliminary calibration for what *might* be safe magnitudes, durations and rates for climate engineering interventions. However, the current imprecision of the concept—due to the inherent complexity of the climate system and GHG-induced climate change—means that "staying within the natural variability of the climate system" is not a particularly useful objective against which to evaluate detailed SWCE system designs.

---

60 NAS-1992, page 435.

61 Ibid.

62 Ibid. Full quote from text: "it seems reasonable to assume that mitigation systems that put dust or aerosols into the atmosphere at altitudes and in quantities that are within the bounds of the natural experiments or of previous experiments would not produce instabilities or effects that had not been produced before."

63 For example, see: Wigley-2006. Quote from text: "We know, for example, that the Mount Pinatubo eruption (June 1991) caused detectable short-term cooling but did not seriously disrupt the climate system. Deliberately adding aerosols or aerosol precursors to the stratosphere, so that the loading is similar to the maximum loading from the Mount Pinatubo Eruption, should therefore present minimal climate risks."

64 NAS-1992, page 435.

## *2.3 Engineering Control Variables for Stratospheric Aerosols*

Any SWCE concept will have a limited number of control variables ($\phi_1$, $\phi_2$, $\phi_3$…) that can be technically "engineered" to optimize against the design objectives. These variables are the primary means through which all climate parameter responses ($\Delta_a$, $\Delta_b$, $\Delta_c$…) for a specific SWCE intervention can be tailored. Both fundamental and practical technical constraints will inevitably limit the range and precision of the engineering possible with each control variable.

Moreover, the *response functions* mapping the climatic impacts to these variables (*e.g.* $\Delta_{\text{Global Average Temperature}} = f[\phi_1, \phi_2, \phi_3\ldots]$) must incorporate the behavior of numerous interdependent climatic processes over a wide range of spatial and temporal scales. Complex response delays, feedbacks, nonlinearities and instabilities are all possible, many of which could lie outside the observed behavior of the climate system, leading to their being poorly understood or even unknown.

The control variables and response functions are therefore central foci of any comprehensive SWCE research agenda.

As control variables will differ significantly among concepts, we focus (as in Part 3) on the illustrative example of aerosol-based stratospheric albedo modification.[65] For this set of SWCE concepts, we have identified five core control variables:

(1) Material Composition(s) of Aerosol Particles

(2) Size(s) (and possibly shapes) of Aerosol Particles

(3) Amount of Aerosols Dispersed

(4) Geographical and Vertical Location(s) of Aerosol Dispersion

(5) Temporal Sequencing of Aerosol Dispersion

All other relevant aspects of an intervention, from intermediate characteristics (*e.g.* radiative properties and stratospheric lifetimes of the aerosol particles) to the final climate responses (*e.g.* variations in atmospheric composition due to chemical reactions of the aerosols and the spatial and temporal distributions of surface insolation), will depend upon this set of core variables. In some cases, the response functions are well understood—for example, Box 2.3.1.1 describes the dependence of spectral scattering efficiency upon particle size for a specified material composition. However, for many other cases, considerably more complex and poorly understood chemical and physical processes are involved—for example, the change in plant growth attributed to enhanced diffuse illumination post-Pinatubo.

The control variables are also subject to constraints and couplings that hinder their independent manipulation. For example, Box 2.3.1.2 illustrates how fundamental physical and chemical properties and process considerations can limit the choice of aerosol material, and how this choice subsequently constrains the possible range of particle sizes. Practical technical issues related to methods of aerosol dispersion in the stratosphere (discussed in Section 3.1.1) add further constraints and couplings to these variables.

---

65 Based on the observed climatic response to volcanoes (see Box 2.1.2.3), Budyko-1977 originally proposed that the intentional injection of sulfate aerosols into the stratosphere could be used to cool the Earth. Since that time, others have proposed using other aerosol materials or small solid particles to accomplish the same objective. For recent reviews, see Rasch-2008b and Lacis-2008.

Even given such constraints, there is a large range of possible interventions, and an even larger number of response functions that could be considered.

As the shortwave radiation (SWR) incident on Earth's surface is the climate parameter most directly affected by SWCE, its spatially and temporally dependent response function will be a foremost consideration (*i.e.* $\Delta_{SWR}(x,y,z,t) = f[\phi_1, \phi_2, \phi_3 \ldots]$), as will the response functions for temperature and precipitation. Indeed, the ongoing preliminary research into SWCE[66] has had this pragmatic basic focus. But as illustrated by the discussion in Section 2.1, the number of climate parameters explored must be expanded to span a range of spatial and temporal scales.

An important research step is therefore the identification of the appropriate set of climate parameters for which response functions should be explored—this is discussed further in Section 3.1.2.

A practical research agenda will begin by exploring these variables within or near the ranges observed for natural analogues to SWCE (*i.e.* sulfate aerosol interventions), thereby providing the greatest possible understanding should a near-term deployment be determined to be necessary. However broader exploration of the viable range for these variables should also be considered. For example, as highlighted in Box 2.3.1.2 below, there are already some proposals for aerosols composed of materials not naturally observed in the stratosphere, and a variety of entirely unexplored possibilities remain. Aerosols engineered to have characteristics not found in naturally-occurring aerosols—for example, scattering cross-sections tailored to wavelengths longer than those used in photosynthesis—could offer unique capabilities, along with unique risks. Another example would be exploring the potential utility of variable spatial and temporal distribution for multiple aerosol materials with different stratospheric lifetimes and radiative properties, in order to provide more regional or even local climatic influence. Knowledge developed from such broader explorations would provide greater flexibility in SWCE system design. The comprehensive research agenda outlined in Part 3 therefore incorporates a balance of both practically-oriented research into basic sulfate-aerosol interventions and exploratory research into novel aerosol material compositions, structures and distribution alternatives.

---

[66] Stratospheric sulfate aerosols, with sizes and in concentrations similar to those observed for past volcanic eruptions, and their impacts temperature and precipitation, have been the primary focus of such studies to date. See Section 2.1.1 for more detail.

**Box 2.3.1.1 Correlation Between Aerosol Particle Size and Scattering Efficiency**

As examined more extensively in the recent publications of Rasch *et al* (2008) and Lacis *et al* (2008),[67] aerosol particle size has a large influence on aerosol scattering of solar radiation. In particular, the *mass scattering efficiency* ($\varepsilon$) of aerosol particles varies significantly with both the diameter of particle ($d$) and the wavelength of the light ($\lambda$) being scattered.

Scattering from homogeneous spherical particles is described as Mie scattering, and data are widely available.[68] For small particles ($d \ll \lambda$) the mass scattering efficiency ($\varepsilon$) scales with the cube of the diameter ($\sim d^3/\lambda^4$), making very small particles ineffective. For large particles ($d \geq \lambda$), the mass scattering efficiency scales with the inverse of the diameter ($\sim d^{-1}$), leading to large particles being inefficient in terms of scattering per unit mass. For materially-uniform spherical particles, basic scattering theory demonstrates that the most mass-efficient scatterers have radii of about 0.1 of the wavelength of the scattered radiation. For the solar spectrum, peaking on the red side of the visible spectrum, this means particles with radii of ~1000Å (or 0.1μm.)

Optimizing the size of injected stratospheric aerosol particles to the point of greatest mass efficiency would minimize aerosol mass that must be lofted into the stratosphere. However, this high mass efficiency comes at the price of increasing the amount of large angle forward scattering from the aerosol particles. This means that the most mass efficient particles for reflecting solar radiation back to space would also produce the largest conversion of direct to diffuse sunlight. As discussed in Box 2.1.2.3, the Mount Pinatubo eruption temporarily converted a significant fraction of incident sunlight from direct to diffuse radiation, dramatically increasing (by ~300%) the total amount of clear sky diffuse sunlight reaching Earth's surface, which is believed to have had a notable impact on plant productivity.[69] Whether increased diffuse sunlight is an important consideration remains at present largely unexplored.

Aerosol particles that go beyond materially uniform spheres—such as layered material spheres or non-spherical particles—might also be possible.[70] In particular, some designs may provide highly mass-efficient reflection of solar radiation while making little contribution to diffuse light reaching the Earth's surface. As the radiative scattering properties of such designs are notably different from uniform spheres, case-by-case evaluation of their mass efficiency and impact on diffuse radiation are necessary.

---

67 Teller-1997; Rasch-2008a; Lacis-2008.

68 The asymptotic forms of the scattering cross-section are straightforward. For diameter $d \ll \lambda$, the cross-section ($\sigma$) is just the Rayleigh cross-section, where $\sigma \propto d^6 / \lambda^4$. This leads to a mass efficiency $\varepsilon \equiv \sigma/(\pi\rho d^3/6) \propto d^3 / \lambda^4$. We have retained the wavelength dependence because it is also strong; particles that are effective in scattering blue light are an order of magnitude less effective in scattering red light (hence the blue sky, or blue diesel exhaust). For $d \geq \lambda$, the cross-section $\sigma \propto d^2$, so that $\varepsilon \propto d^{-1}$.

69 Gu-2003.

70 For a discussion of specific examples, please see: Keith-2000 and Lacis-2008.

**Box 2.3.1.2 Aerosol Material Considerations for Stratospheric Albedo Modification**

For stratospheric albedo modification, desirable properties of engineered aerosols include: *(a)* scattering targeted portions of the solar spectrum with minimal absorption (to avoid excessive stratospheric heating—although it has been suggested that some heating might help reduce ozone loss[71]); *(b)* minimal catalysis of stratospheric ozone depletion; *(c)* minimal mass; *(d)* chemical stability in the atmosphere (even under the influence of the large UV flux and strong oxidants in the stratosphere); *(e)* minimal vapor pressure (to prevent their evaporation); and *(f)* minimal scattering or absorption of thermal infrared radiation in the 8–14µ window. These basic properties hint that oxide elements from the third row of the periodic table might be good material candidates, and similarly eliminate some materials from consideration.[72]

Among the most commonly discussed materials for stratospheric aerosols for SWCE have been hydrated sulfuric acid ($H_2SO_4 \cdot H_2O$), carbon-soot, and aluminium or aluminium oxide. Aerosols of these materials (or a number of other candidates) can be divided into two classes: 1) those that kinetically agglomerate from gas-phase material following oxidation and hydration in the atmosphere (*e.g.* hydrated sulfuric acid); and 2) solid particles (*e.g.* carbon-soot, aluminium or aluminium oxide.) Both classes (and all specific material choices) have advantages, disadvantages and outstanding dispersal challenges that require further investigation before a robust comparative evaluation of their merits can be done.

***Aerosols from Kinetically Agglomerated Gas-Phase Materials***

Sulfuric acid particles are the most often considered aerosols for stratospheric albedo modification based on the considerable body of existing observations of their presence in the atmosphere (from both volcanic activity and burning fuels with sulfur impurities.) Additional dielectric candidates include a number of oxides and hydrated oxides (examples include $Li_2O$, $B(OH)_3$, $Na_2O$, MgO, $Al_2O_3$, $SiO_2$ and $H_3PO_4$), however, there are considerably fewer data available regarding their atmospheric interactions and cycling.

Dispersal of the gas phase precursor materials for these aerosols (such as volatile hydride compounds[73]) could have several advantages. Such a method might facilitate aerosol material dispersion, minimize coagulation by delaying oxidation until after the precursor hydrides are diluted in the stratosphere, and reduce the mass that must be lofted because the oxygen and water could be drawn from the atmosphere. Nonetheless, several significant issues still require research. In particular, the kinetics of precursor oxidation and hydration, agglomeration, and coagulation in the stratosphere are poorly understood. Even for the most studied case of sulfuric acid aerosols, the particle size distributions that would be produced by intentional dispersion methods are not well known. Processes such as the condensation of injected $H_2SO_4$ vapor onto existing particles[74] and the scavenging of smaller particles by larger ones through coagulation or vapor phase transport might lead to large particles that are very mass-inefficient at scattering visible radiation (*see Box 2.3.1.1*), and which might also sediment out of the stratosphere rapidly.[75] Further laboratory, computational and potentially field-experiment research into aerosol formation kinetics are required.

[71] Crutzen-2006.

[72] For example, transition metal oxides are generally strong absorbers of visible light (which could cause significant stratospheric hearting) and second row elements other than boron are either rare and toxic in all chemical forms (*e.g.* beryllium) or are GHGs.

[73] For example, hydrogen sulfide ($H_2S$). Some candidate materials might alternatively be lofted in elemental form and burned in the stratosphere (lithium, sodium, magnesium and white phosphorus).

[74] Driven by the low vapor pressure of sulfuric acid.

[75] Richard Turco (UCLA) and Phillip Rasch (NCAR), personal correspondence.

***Solid Particles***

Another possibility is the fabrication of engineered solid particles, which would be manufactured on the ground and then lofted into the stratosphere;[76] proposals have focused particularly on aluminium and aluminium oxide particles. The resulting high degree of control over particle sizes and shapes could provide a number of advantages, including high mass scattering efficiency[77] and possibly control over the angular scattering function thereby significantly reducing the direct-to-diffuse conversion of solar radiation discussed above in Box 2.3.1.1. Such particles might even be designed to enable photophoretic levitation,[78] which may enable them to be concentrated zonally and further reduce the mass and lofting requirements (due to potentially very long stratospheric lifetimes.)

A crucial challenge facing any solid aerosol particles is dispersion. Electrostatic and van der Waals forces can bind small particles together quite strongly unless they are dispersed in a solvent. However the use of solvent could markedly increase (possibly by orders of magnitude) the mass lofting requirements. Research into monolayer particle coatings and dispersion methods (not yet investigated in this context) would be needed to determine the difficulty and cost of overcoming these challenges. Such research could build on experience with the dispersal of small particles for applications such as paints and coatings. For example, van der Waals attraction might be reduced with fluorinated coatings or with electrostatic forces by using conducting particles.

---

[76] See for example and discussion: Teller-1997; Keith-2000; Keith-2008; Lacis-2008.

[77] For example, Teller-1997 proposed short (~300nm) and fine (thickness and width ~30nm) aluminum wires, coated with aluminum oxide for protection against further oxidation. The required mass of such particles would be only a few percent, at most, of the mass of sulfur required to achieve the same radiative effect.

[78] Keith-2008.

## *2.4 Phases of Research, Development and Deployment for Shortwave Climate Engineering*

Climate engineering science and technology are in their infancy. SWCE investigations to date have been limited to speculation, paper studies, and preliminary climate simulations of uniform SWCE with coarse-resolution models. Targeted and directed investigations across a wide range of subjects—from basic climate science to intervention system engineering—could significantly improve our understanding and reduce uncertainty in the climate response to a given SWCE intervention. Whether the results reveal generic unforeseen hazards that discourage SWCE or highly predictable climatic responses that provide confidence in specific SWCE concepts, such studies would invariably improve our ability to evaluate the risks and benefits of any potential SWCE response to a climate emergency.

As outlined in the Prelude, however, there are perceived risks, both technical and socio-political, associated with pursuing climate engineering research. This study made no attempt to evaluate those risks comprehensively, but we did distinguish between several categories of investigation that present different types and magnitudes of risk:

- *(I) Non-Invasive Laboratory/Computer-Based Research and Related Case Studies*
- *(II) Field Experiments*
- *(III) Monitored Deployment*

These categories can simplistically be viewed as sequential phases, with each subsequent phase presenting greater risks (both technical and socio-political) while providing data of increasing utility for reducing uncertainty (and with *Phases II* and *III* iteratively complementing, not supplanting, ongoing *Phase 1* efforts.) The general characteristics of each phase are discussed throughout this section and summarized in Figure 4.

Valuable analogues are the *(I) pre-clinical testing*, *(II) clinical trials*, and *(III) clinical use* stages of medical drug and device development. Clinical trials and use entail increasing risks to human patients, along with greater understanding of treatment effectiveness and side effects, and the potential improvements in patients' health. Before a decision is taken to escalate from one stage to the next, tradeoffs among the potential risks and benefits must be carefully weighed based upon the results of preceding stages. (However, two obvious but crucial differences from the medical analogy—the lack of "animal-testing" equivalents and the lack of an ensemble of "patients"—render inapplicable many of the standard statistical risk-evaluation and management tools of the medical process.)

A detailed evaluation of the tradeoffs in research progression was beyond the scope of our study, but we indentify below basic technical issues related to the risks for each phase. These phases are also employed in Section 3 to organize the components of the comprehensive research agenda presented for the example of aerosol-based stratospheric albedo modification.

### *2.4.1 Phase I: Non-Invasive Laboratory/Computer-based Research and Analogue Case Studies*

*Phase I* investigations utilize existing understanding and observations of the climate system to non-invasively explore potential climate system responses to SWCE interventions—predominantly using laboratory experimentation and computational modeling. This phase involves no intervention into the climate system, at any scale, and therefore presents no risk of unintended climatic consequences.

Current ongoing computer-based research into SWCE[79] (*Box 2.1.2.2*) and evaluations of natural experiment analogues[80] (*Box 2.1.2.3*) are examples of this category of research. As suggested above in Section 2.1 and described in more detail below in Part 3 (for the specific case of aerosol-based stratospheric albedo modification), very significant scope exists for expanding these research efforts. This includes new research avenues evaluating unintended perturbations in the anthropogenic SWCE forcing of the climate system[81] and a range of laboratory experiments that could further enrich this phase.

Two additional key components of *Phase I* research are the already ongoing processes of climate emergency threat assessment and baseline climate data collection that are part of the international effort to understand climate change. These provide essential data for any comprehensive net risk assessment relating to climate engineering research or interventions.

### *2.4.2 Phase II: Field Experiments*

*Phase II* investigations incorporate intentional interventions into the climate system that are limited in duration, magnitude, and/or spatial range. The primary goal of such interventions would be to investigate SWCE concept efficacy rather than to counter specific aspects of GHG-induced climate change (though the potential for such results may be a secondary consideration in the experimental design.) Regardless of the limits and precautions put in place, such experiments at regional or larger scales would inevitably involve some risk of undesired climatic consequences. On the other hand, direct empirical testing provides the only incontrovertible method for establishing the efficacy of SWCE concepts.

Using the results of *Phase I* investigations, any *Phase II* experiments should be designed to strike an appropriate balance between two conflicting objectives: (a) minimizing the possibility of negative climatic consequences (expected or unexpected); and (b) maximizing the utility of collected data for increasing understanding. The tradeoffs between these two goals should be carefully evaluated during the experimental design periods, along with an assessment of any associated socio-political risks.

When assessing these tradeoffs, it should be recognized that the understanding gained from field experiments does not necessarily increase commensurately with field experiment scale. A range of signal-to-noise limitations associated with natural climate variability and response timescales can severely constrain the conclusions that might be drawn from any experiment. These limitations are discussed in greater detail in Section 3.2 in the context of designing field-tests for aerosol-based stratospheric albedo modification concepts.

### *2.4.3 Phase III: Monitored Deployment*

*Phase III* incorporates the deployment and monitoring of an intentional intervention into the climate system, with the primary goal of inducing a desired climatic response. Deployment of an SWCE system will inevitably involve additional risks beyond those of *Phase II* field experiments due to: (1) the incomplete understanding of climatic consequences that can be gained even from field experiments; and (2) the possibility of unanticipated climatic consequences becoming apparent only well after an intervention has begun (particularly for climate components with long equilibration times, such as large ecological systems.)

---

[79] Bala-2008; Matthews-2007; Bala-2007; Bala-2006; Govindasamy-2003; Govindasamy-2002; Govindasamy-2000.

[80] Trenberth-2007.

[81] Examples include the temporary Chinese shutdown of factories for the 2008 Olympics or the grounding of all flights in North America immediately after September 11, 2001.

The technical risks at this stage will also depend on the quantify and quality of *Phase I* and *II* research already done, as well as on the state of the climate system at the time of deployment. Any decision to deploy will involve trade-offs between the potential risk reduction by further research, and the mounting real or perceived risks from growing climate stresses. Such a decision is also not binary, with a continuous spectrum of options existing between gradual deployment (allowing significant time for further research and optimization, including careful monitoring during low level deployment), and immediate rapid deployment. As discussed in Section 2.1.2, this decision should be influenced by the state of the climate system at the time, with different climate emergencies calling for different deployment timescales and magnitudes.

Taken together, these deployment-related risks emphasize the importance of extensive and ongoing monitoring of the climate system in this phase, as well as the value of scalable deployment and rapid disengagement capabilities for any SWCE intervention.

## *2.5 Maintaining Interventions and the Risks of Stopping Once Started*

Prior to deploying any SWCE intervention, the long-term challenges associated with its maintenance (*Phase IV*) and eventual disengagement (*Phase V*) should be rigorously considered. The specific details of the system design and intervention target(s) will significantly define these challenges, but some general observations can still be made at this point. The general characteristics of these additional phases are also summarized below in Figure 4.

### *2.5.1 Phase IV: Intervention Maintenance*

One-time interventions, analogous to artificial volcanoes, would have the lowest maintenance requirements. As observed for Mt Pinatubo, the major climatic impacts of such interventions are likely to be short-term (*see Box 2.1.2.3.*) Nonetheless, a strong transient intervention might be useful for delaying the arrival of particularly damaging GHG-induced climatic changes, thereby allowing some limited additional time for natural and societal adaptations.

After an initial deployment period—which could entail anything ranging from a slow and gradual ramp-up to a rapid and strong transient intervention—most envisioned SWCE interventions would settle into a long-term steady-state phase. During this stage (*Phase IV*), ongoing adjustment of the control variables (*see Section 2.3*) in response to continuous climate observations would be essential to maintain and optimize the intervention's climatic impacts. The development of a dynamic multivariate control-system—incorporating robust monitoring of climate parameters, maximal intervention flexibility and intervention stability—is therefore an important component of SWCE research. Control-system design should pay particular attention to the likelihood of various climate parameter responses including delays, feedbacks, nonlinearities and instabilities across widely ranging temporal and spatial scales.

The GHG emissions scenario occurring in parallel to any steady-state intervention will also be a central consideration in this stage. The potential targets for an intervention vary significantly—from globally compensating (to the degree possible) all anthropogenic-GHG climate impacts into the future to inducing specifically desired regional climatic responses (*see Section 2.1.2.*) But for each, changing concentrations of GHGs define the background context in which any target will be maintained. For example, the expected fidelity with which SWCE could offset climatic impacts decreases with increasing GHG concentrations (*see Section 2.1.1.*) Consequently, strategic management of GHG emissions must be considered a central component in the maintenance of any SWCE intervention.

### *2.5.2 Phase V: Disengagement*

Should a steady-state SWCE intervention ever be disengaged intentionally (*Phase V*), risk-tradeoffs similar to those for the deployment stage (*discussed above for Phase III, Section 2.4.3*) must be considered, and those risks will depend on the rate of disengagement. If serious harmful side effects are discovered, rapid disengagement might be called for.

However the risks associated with the transient climatic "rebound" to such a rapid disengagement must also be considered carefully. For example, simulations by Matthews and Caldeira (2007) and Brovkin (2008)[82] have suggested that sudden disengagement could cause short-term rates of global temperature change more than an order of magnitude greater than those observed today. The magnitude of these rebound effects will scale—though not necessarily linearly—with the magnitude of the SWCE intervention, once again revealing increasing technical risks associated with attempting to offset higher GHG concentrations. Any intentional disengagement must therefore balance the risks of a rapid disengagement against those of prolonging any realized (or avoiding any imminent) harmful side effects through gradual disengagement.

Technical or socio-political SWCE system failures could also lead to unintentional disengagement. For permanent failures, the consequences would be the same as for rapid intentional disengagement. But temporary system disruptions are also possible, and these could induce transient oscillations in the climate system with unique impacts related to the magnitude and duration of the disruption. As disruptions of varying character and scale are common in comparably large and complex technical and socio-political systems, any SWCE control-system should be designed with such possibilities explicitly in mind. In particular, redundancy mechanisms and issues related to controlled reengagement after a disruption should be investigated.

The possibility of counter-climate engineering must also be considered. Should any sufficiently large global actor decide, after an SWCE intervention had begun, that it was disadvantaged by the induced climatic changes, it could be technically possible for it to develop and deploy *countermeasures* to achieve results similar to disengagement. For example, the deliberate injection of short-lived fluorocarbon GHGs might rapidly offset the regional or global cooling effects of a SWCE intervention. However the complex and global climatic consequences of such countermeasure strategies could be disastrous. Though international socio-political issues are beyond the scope of this study, the technical possibility of such countermeasures emphasizes the importance of carefully considering the international relations and governance implications of climate engineering.

Should an SWCE intervention be deemed a necessary response to a climate emergency, the ideal case would be that an SWCE intervention successfully achieves its target, that the need for it gradually diminishes, and that it is thus gradually phased out. Such a scenario could be driven by eventual reductions in GHG concentrations or by acceptance of gradual societal and ecological adaptation to higher concentrations. However it is also possible that such a scenario does not manifest itself, and that once engaged, the maintenance of a SWCE system becomes a permanent bequest to future generations.

[82] Matthews-2007; Brovkin-2008.

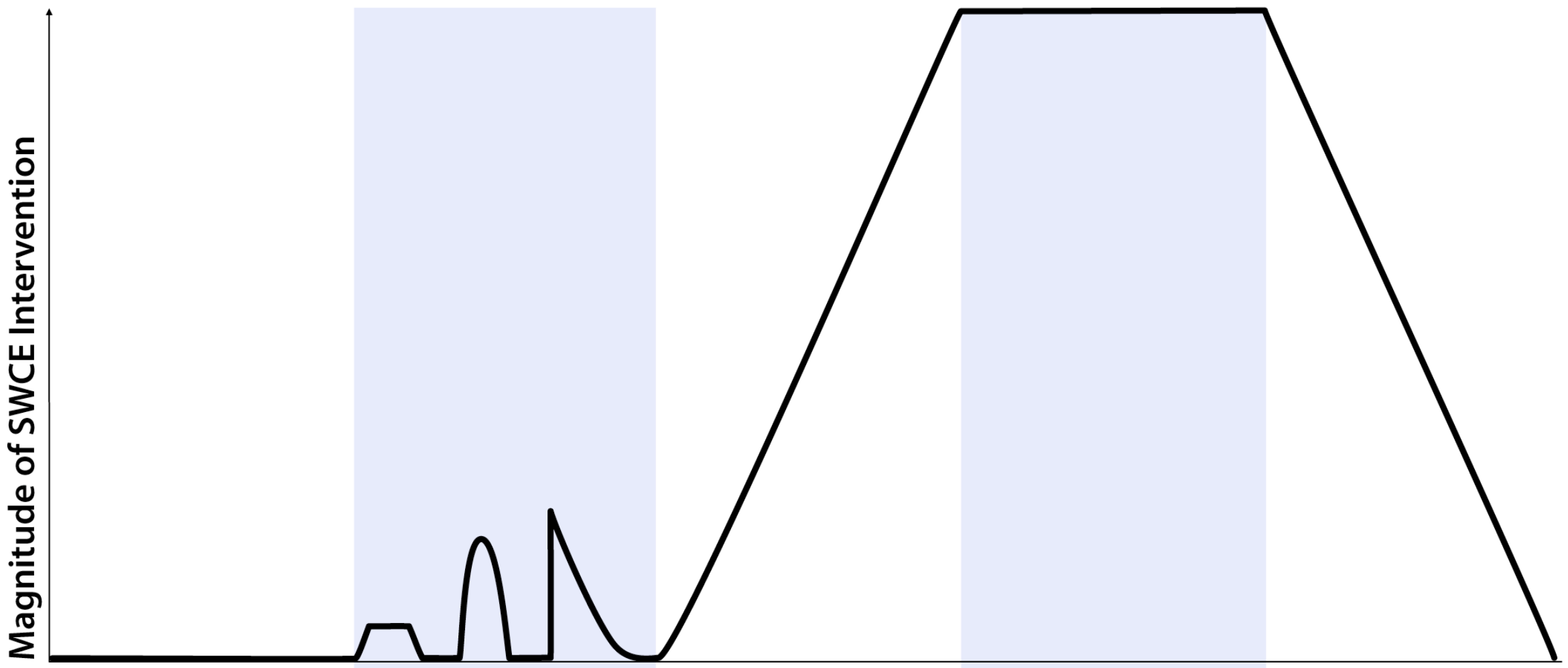

### Phase I
**Non-Invasive Research**

*Objective* Utilize existing understanding of climate system to evaluate SWCE concepts

*Components*
- Laboratory experiments
- Computer modeling
- Analogue case studies

*Important Characteristics*
- No direct climatic impacts or associated risks
- Limited learning potential without some iterative empirical testing (Phase II research)
- Existing evaluations of SWCE concepts are Phase I research (*see Boxes 2.1.1.2 and 2.1.1.3*)

### Phase II
**Field Experiments**

*Objective* Develop understanding of potential SWCE concepts through limited scale intervention experiments

*Components*
- SWCE interventions limited in duration, magnitude and/or spatial range
- Climatic impact monitoring

*Associated Issues*
- Climatic impacts and associated risks increase with scale of field test (though not necessarily increasing proportionally)
- Signal-to-noise limits associated with natural variability and temporal/-spatial delays in climatic responses constrain potential learning from field-tests (*see Section 3.2 for discussion*)
- Empirical data from tests could iteratively improve Phase I research

### Phase III
**Monitored Deployment**

*Objective* Achieve a desired state of the climate system

*Components*
- Deployment of a full-scale SWCE intervention (possible ramp-up timescale between gradual and immediate)
- Climatic impact monitoring
- Development of a control-system for the SWCE intervention system (based on monitored climate parameters)

*Associated Issues*
- Increasing climatic impacts and risks with increasing SWCE intervention scale (though not necessarily increasing proportionally)
- Gradual deployment could be similar to Phase II research, allowing time for testing iterative improvement of Phase I research

### Phase IV
**Steady-State Intervention**

*Objective* Maintenance of a desired state of the climate system

*Components*
- Maintenance and improvement a full-scale SWCE intervention
- Climatic impact monitoring
- Improvement of the SWCE intervention control-system

*Associated Issues*
- Long-term SWCE could generate cumulative climate impacts and risks unobserved in field-tests (Phase II) or initial deployment (Phase III)
- Increasing atmospheric GHG concentrations will have a separate impact on the climate system that must be incorporated into any intervention control-system

### Phase V
**Disengagement**

*Motivations for Intentional Disengagement*
- Reduction in need due to successful achievement of intervention target (e.g. mitigation reduces GHG levels and SWCE to prevent warming no longer required)
- Discovery of harmful side effects of the intervention

*Possible Causes of Unintentional Disengagement*
- Technical or socio-political system failure
- Counter-climate engineering or countermeasures

*Associated Issues*
- Potential for climatic parameter rebound effects
- Severity of climate parameter rebound will increase with intervention scale and disengagement rate

***Figure 4*** Phases of SWCE research, development and deployment

Conceptual representation of the magnitude and temporal character (*e.g. short-term or intermittent, increasing/decreasing or steady-state*) interventions in the climate system associated with the various stages of SWCE research, development and deployment.

# 3 Reducing the Uncertainties for Stratospheric Aerosols

## *3.1 A Comprehensive Research Agenda*

The injection of stratospheric aerosols to cool the globe rapidly is among the most frequently considered SWCE interventions, both because of its similarity to observed natural events (*see Box 2.1.2.3*) and because it is seen to be technologically achievable with low implementation costs (*see Section 3.1.1*.) Though not necessarily an appropriate response for *all* possible climate emergencies (*see Section 2.1.2*), this family of interventions has the potential to offset many climatic impacts associated with GHG-induced warming (*see Section 2.1.1*.) But as highlighted repeatedly in this report, technical investigation of even this concept has been limited to basic theoretic studies and idealized exploratory simulations.

Our study therefore concentrated on what could be done with a decade of concerted research to maximally reduce uncertainty in possible stratospheric aerosol interventions. Significant improvements in our understanding are possible in this timeframe, allowing for much more confident risk-effectiveness assessments should a climate emergency arise—but many outstanding research questions and uncertainties would inevitably remain after even a decade of systematic and coordinated research.

The research agenda we outline is divided into three major streams, each focusing on a central issue identified in Chapter 2. The first ("Engineering" or "E") stream focuses on the technical issues associated with design and deployment of a stratospheric aerosol SWCE system. Following from the end of Section 2.3, this stream incorporates a branch considering practical deployment technologies for "nature-imitating" sulfate aerosols, along with a branch exploring the design space of engineered aerosol particles. The second ("Climate Science" or "CS") stream also builds upon Section 2.3, concentrating on understanding the response of the climate system to a stratospheric aerosol SWCE intervention. Once again, there are two branches, one for sulfate aerosols and the other for exploratory studies of the climate response to interventions with engineered aerosols. The crucial third ("Climate Monitoring" or "CM") stream focuses on the development of broad climate monitoring capabilities to detect the impact of an SWCE intervention for both evaluation and control purposes (*discussed in Sections 2.4.3 and 2.5*.)

The three streams are described respectively in Sections 3.1.1, 3.1.2, and 3.1.3 below. For each, current understanding and ongoing research are summarized, and core research questions are identified. Specific research avenues, methods, and tools for addressing these questions are then described, with each component being identified as either *Phase I* (non-intervention) or *Phase II* (field-testing) SWCE research (*see Section 2.4 and Figure 4*.) For the CS and CM streams, strategies for building upon ongoing efforts to monitor and understand climate are particularly important. The core research questions for these streams are summarized in Figure 5, which outlines a rough timeline for a comprehensive decade-long research program integrating these three streams.

As discussed in Section 1.3, the technical research agenda we outline would maximally reduce uncertainty in stratospheric aerosol SWCE over the next decade. This is not a recommendation that this program of research *should* be pursued (*see the Prelude to this report*.) However, we note that many of the CS and CM research activities required for improving our understanding of stratospheric aerosol SWCE would also significantly reduce uncertainties related to natural and anthropogenic radiative forcing and ongoing climate change.

While some such benefits of aerosol field experiments are discussed briefly through this section, a comprehensive evaluation was beyond the scope of this study. A future study specifically examining these potential benefits is well warranted.

### *3.1.1 E Stream Research: Engineering Deployment Technologies and Aerosol Particles*

SWCE with stratospheric aerosols minimally requires: (1) the ability to produce bulk quantities of aerosol particles with appropriate radiative properties; and (2) the ability to disperse and maintain a concentration of those aerosol particles in the stratosphere, ideally with spatial and temporal control of their distribution.

The 1992 National Academy of Science (NAS) review estimated that a basic stratospheric aerosol delivery system:

> "appear[ed] feasible, economical, and capable of [offsetting] as much $CO_2$ equivalent per year as we care to pay for… [and] could probably be put into full effect within a year or two of a decision to do so."[83]

Although we did not explore deployment costs in any detail, our order-of-magnitude cost estimates for utilizing aircraft, guns or rockets for aerosol lofting (between ~$10B and ~$30B per year to loft $10^9$ kg per year,[84] as calculated in Appendix I) agrees roughly with the NAS and other previous cost estimates.[85] Our technical review of aerosol deployment options (*see Box 3.1.1.1 and Appendix I*) confirms the feasibility of a rapid deployment of sulfate aerosol SWCE intervention. Interventions with more highly engineered particles are also likely feasible, but would need to be evaluated on a case-by-case basis. However, we emphasize that rapid deployment of SWCE prior to comprehensive CS and CM research would induce a climate response laden with uncertainty and risk. These uncertainties and risks would be even greater for engineered particles that are significantly different from those in the volcanic "natural experiments" described above in Box 2.1.2.3.

To our knowledge, only a handful of investigations since the NAS review have evaluated stratospheric aerosol deployment technologies or non-sulfate aerosols.[86] Numerous methods for lofting the necessary mass of aerosol or precursor material to the stratosphere have been discussed, and several prominent methods are reviewed in Box 3.1.1.1 (and further analyzed in Appendix I.) However, a number of important considerations have been only cursorily examined; for example, the potential for rapid aerosol agglomeration and sedimentation even after successful dispersion (*as discussed in Box 3.1.1.2.*)[87] The discussion in Box 2.3.1.2 shows that research into potential aerosol materials has been similarly preliminary. These examples illustrate that research into aerosol and deployment technologies remains in its infancy.

---

83 NAS-1992, page 459.

84 The stratospheric sulfate aerosol loading necessary to offset a doubling of $CO_2$ is currently estimated to be between $1.5\times10^9$ kg (1.5 Tg) and $5\times10^9$ kg (5 Tg). Assuming stratospheric aerosol lifetimes on the order of one year, approximately this mass of aerosol (or aerosol precursor) material would need to be lofted each year. A nominal $10^9$ kg/yr (1 Tg) lofting rate is utilized for basic calculations in Appendix I.

85 For a brief summary of cost estimates and references, see: Barrett-2008.

86 Teller-1997; Teller-2002; Crutzen-2006; Rasch-2008a; Lacis-2008; Keith-2008.

87 The year-long effects of historic volcanic injection suggests some empirical bounds for these processes, at least for similar material (sulfate) aerosol particles (though the similarity of man-made sulfate aerosols to volcanic sulfate aerosols remains questionable).

The five *control variables* for stratospheric aerosol interventions given in Section 2.3 essentially define the core research questions of the E stream. Given the decade of research that we have specified in this study, one branch of investigation should focus on sulfate aerosols, with the following key questions:

(1) What are the possible and optimal materials and dispersion methods to facilitate stratospheric loading with aerosols of appropriate size and composition?

- *Some of the issues that should be explored include the chemical kinetics of oxidation of gaseous hydride precursors and the kinetics of aerosol aggregation.*

(2) To what minimal altitudes must (and optimal altitudes should) the materials be lofted at different latitudes? How do the lofting altitude, location, and temporal sequencing of aerosol injection determine the temporal and spatial distribution of aerosols in the stratosphere across the globe?

- *Necessary studies include global wind fields (including up- and down-welling), turbulent diffusion (particularly vertical), and agglomeration and sedimentation processes.*

(3) What are the possible and optimal lofting methods given different mass (or volume), altitudinal, and spatial injection requirements?

- *Some of the issues to be explored for various lofting methods are discussed in Box 3.1.1.1 and Appendix I.*

Even within a decade-long research program, the unique capabilities offered by engineered aerosols should also be explored. As discussed briefly at the end of Section 2.3 and in Box 2.3.1.2, aerosol particles could be created from many materials. Moreover, for many materials, control of particle size and structure might be possible (*e.g.* non-spherical, multi-layered, or coated particles.) For any engineered particle design, a fourth question supplementing the above three must be:

(4) What are the radiative impacts and environmental interactions of the engineered aerosol?

The answers to this question for sulfate aerosols are more clearly understood than for other materials because of natural experiments (*see Box 2.1.2.3*) and current tropospheric sulfate aerosol pollution. However, even for sulfate aerosols, potential differences between natural and artificial aerosols, and stratospheric and tropospheric aerosol impacts and interactions, significant uncertainties still remain and require investigation.

*Phase I* laboratory and computer-based experiments to address each of these questions can be straightforwardly envisioned. Chambers that reproduce the stratospheric environment (pressure, temperature, chemical constituents, solar illumination) could be useful to determine chemical and physical kinetics. Simulations of stratospheric transport would provide some understanding of minimal altitudes and global distributions (modeling is discussed in greater detail in Section 3.1.2.) A range of laboratory experiments could also address many questions related to materials for, and durability of, guns, balloons and chimneys (*see Appendix I for details*.)

One particular class of "field experiment" would also qualify as *Phase I* since they do not involve an intervention of any scale into the climate system. Experiments that test lofting technologies without aerosol deployment fall into this category. The dispersion of environmentally-inert tracer particles with negligible radiative impacts to probe stratospheric transport are at the boundary between *Phase I* and *Phase II* tests, but would still qualify as *Phase I* if negligible environmental impact could be demonstrated *a priori*.

This spectrum of *Phase I* research could dramatically reduce current uncertainty, but fully addressing each the relevant questions will minimally require low-level field experiments that have some potential for detectable impact on the climate system (i.e. *Phase II* research.) In particular, *in situ* validation of scaled laboratory experiments and simulations for Questions 1 and 2 will be necessary, together with demonstration of scalability to deployment capacity for lofting methods in Question 3. Such field experiments are considered further in Section 3.2.

**Box 3.1.1.1 Lofting Aerosol Material to the Stratosphere**

A number of methods of lofting aerosol or aerosol precursor materials to stratospheric altitudes can be considered. Here (*and also in Appendix I*), we review a set of the most prominent alternatives—aircraft, guns, rockets, suspended chimneys, and balloons. The feasibility of each depends upon the material being lofted and the means of dispersion, and are not independent of the issues discussed in Box 2.3.1.2 and Box 3.1.1.2. We caution that our preliminary evaluations are no substitute for proper engineering design studies of these options.

The most important consideration—and the largest current source of uncertainty—affecting the choice of lofting mechanism is the height to which materials must be lofted. Significant fundamental questions regarding aerosol transport in the stratosphere need to be addressed before any minimal lofting height for stratospheric dispersal could be confidently established. The presence of equatorial stratospheric upwelling suggests that altitudes of 20 km may be sufficient—any material injected there or higher would be transported vertically by the equatorial upwelling and then distributed throughout the stratosphere. This suggests that existing aircraft may well be suitable and sufficient for the purpose.

However, the time scale and geographic distribution resulting from longitudinal spreading of materials over the rest of the globe have yet to be understood. For greater distributional control or targeted interventions (*e.g.* over the arctic), injection at high latitudes or over the poles may be desirable, but a significantly higher lofting altitude (greater than 30 km) may be necessary to compensate for the down-welling in the polar stratosphere. Detailed transport calculations, founded on empirical velocity fields and turbulent transport data, combined with tracer field experiments, would be required to fully address this fundamental question. Furthermore, the use of alternative aerosol materials—such as particles engineered to be lofted from lower altitudes into the stratosphere by photophoretic lift[88]—could significantly influence the lofting requirements.

In Appendix I we examine in greater detail the most prominent options for stratospheric lofting. The rough estimates and calculations there are not a substitute for proper engineering analyses, but they reveal the very limited state of current knowledge about several of the frequently discussed concepts. Of those considered, two lofting concepts capable of reaching almost any stratospheric altitude—rockets and guns—are technically mature, and would require only engineering development. Chimneys would require extensive research and development and it is difficult to estimate their cost.

[88] Keith-2008.

**Box 3.1.1.2 Potential for Aerosol Agglomeration in the Stratosphere**

Agglomeration, even after successful dispersion, can still reduce the residence time of aerosol particles in the stratosphere. Particles that agglomerate may become too large to be mass-efficient scatterers, and they may become so large that gravitational settling removes them from the stratosphere too rapidly.

The Brownian mean free path (*mfp*) of a spherical particle is:

$$mfp_{particle} = \frac{4}{3C_D}\frac{\rho}{\rho_a}r \approx 5 \times 10^4 r$$

where $\rho$ is the particle density (which we have taken to be ~1.8 gm/cm$^3$, which is valid for sulfuric acid, and close for many of the alternative materials considered), $r$ the particle radius, $\rho_a$ the density of the surrounding air, and $C_D$ (~1) is the drag coefficient. This result is valid in Knudsen flow, in which the mean free path of the air molecules is much greater than the particle radius ($r$), which is the case in the stratosphere. When $mfp_{particle}$ is much greater than the particle radius (as is always valid for such small particles), the coagulation time of uncharged monodisperse particles is simply:

$$t_{coag} = \frac{2}{nK}$$

where $n$ is the particle number density and the coagulation coefficient ($K$) is defined by:

$$K = A\sqrt{\frac{24\pi k_B Tr}{\rho}} \approx 4 \times 10^{-9} cm^3 / sec$$

where $A$ is the sticking coefficient and $r$ = 1000 Å. These results are similar to those of Rasch *et al.* (2008.)[89]

Agglomeration may be a particularly acute problem in the initial stages of dispersion when the particle density is high. One way to avoid this problem might to introduce the cation of the scattering material in the form of its volatile hydride (*as also discussed in Box 2.3.1.2*) so that aerosols are not formed (because the oxidation kinetics are not instantaneous) until the vapor has dispersed to low density. On the other hand, if oxidation is too slow, the individual oxide molecules will not form aerosols large enough to be effective scatterers.

Increasing the Earth's albedo by 3% with particles with radius of 1000 Å and cross-section of ~$3\times10^{-10}$ cm$^2$ implies a column density ~$10^8$ /cm$^2$. Distributed over a column 10 km in height, the particle number density ($n$) is ~$10^2$ cm$^{-3}$. This yields a coagulation time ($t_{coag}$) of ~$5\times10^6/A$ sec, or $2/A$ months. Unless $A$ is less than ~0.1, than this coagulation time is prohibitively short. However, the observed persistence of volcanic aerosols (*see Box 2.1.2.3*) implies that this is the case for at least some aerosols.

Methods for inducing aerosols to have values of $A$ less than ~0.1 need to be explored, and could include monolayer coatings or electrostatic repulsion. Either of these options could be a deliberately engineered feature of the dispersed particle, or unintentionally acquired in the stratosphere by interaction with stratospheric gases (coatings) or ultraviolet radiation (charge.) In all cases, both equilibrium and kinetic studies are necessary to explore the utility of these (and other) methods for preventing agglomeration.

[89] Rasch-2008b.

### *3.1.2 CS Stream Research: Climate Science*

At the heart of research into stratospheric aerosol interventions will be questions surrounding their potential climate impacts. Could proposed interventions produce desirable outcomes across all regions and time scales? How much of GHG-induced climate change could they offset? What unintended climate impacts might they produce? Adequately answering these questions—and a myriad of related ones—will require a research effort much larger and deeper than required for the E Stream (*i.e.* aerosols, lofting, and dispersion.)

A review of current understanding of the climate system is far beyond the scope of this report—readers are referred to IPCC AR4 for the most recent comprehensive review.[90] Section 2.1 above provides a succinct review of our current understanding of the potential impact of SWCE interventions on the climate system, illustrating the preliminary state of current research in this area. The research agenda proposed here for this stream focuses particularly on understanding the coupling between stratospheric aerosol interventions and a range of climatic parameters.

Much of the research proposed here would significantly enhance our current understanding of the climate system, and thus of ongoing climate change. A few explicit examples are provided below, but a detailed exploration of such benefits is beyond the scope of this study. We recommend a subsequent study focused on this topic as an initial near-term research objective for this stream—the potential benefits for our understanding of current anthropogenic climate forcing could influence societal decisions about whether parts of this proposed research agenda *should* be pursued.

As noted in Section 2.3, the first practical step in this research stream will be identifying the subset of climatic parameters deemed most important for impact evaluation and monitoring. Though some parameters are easily identified (*e.g.* SW radiative flux at the top of the atmosphere and the surface, global and regional mean surface temperature, global and regional mean precipitation, *etc.*), ensuring the selection of *all* relevant climate parameters (*e.g.,* ecological or atmospheric chemistry parameters) is non-trivial.

For both the CM and CS research streams, we divide the climate parameters into four broad categories: (1) Radiative; (2) Geophysical; (3) Geochemical; and (4) Ecological. Initial research within each of these categories must identify the most appropriate parameters to best characterize the climatic impacts of a stratospheric aerosol intervention. Moreover, these parameters should span the complete range of relevant spatial (local to global) and temporal (daily to centennial) scales, and incorporate measures of the natural patterns of climate variability. Finally—but very importantly—particular attention must be focused on seeking potential *unintended* climate parameter impacts. Completely unforeseeable impacts ("unknown unknowns") will always remain possible. However, many currently unanticipated impacts might be identified through expansive and thoughtful exploration of these categories.

Once key parameters in each category are identified, the second stage of this research stream would aim to define the *response functions* of these parameters to a stratospheric aerosol intervention. It is particularly useful to define these response functions in terms of the *control variables* ($\phi_j$) for such an intervention, as introduced in Section 2.3 (*e.g.* $\Delta_{\text{Global Average Temperature}} = f[\phi_1, \phi_2, \phi_3 \ldots]$.) As noted there, many response functions will be non-local (both spatially and temporally) and complex, potentially including poorly-understood processes, non-linearities, and instabilities.

---

[90] IPCC-2007b.

Consequently, even the most extensive decade-long climate research program will likely leave substantial uncertainty in many of the components in these response functions. Sensitivity analyses exploring the varying climatic impacts at the limits of uncertainty must therefore be a central aspect of research.

The final stage of research in this stream would be the estimation of risks to societal and ecological systems posed by the various climate parameter response functions. The development of such a net risk assessment methodology should begin early, and be structured to integrate continually improving understanding of the important climate parameters and their response functions. This methodology must also address the issue of evaluating risks posed by potential climate emergencies (already the focus of some on-going climate research), and the corresponding palliative utility of a stratospheric aerosol intervention.

The core research questions for the CS stream are thus:

(1) What are the climate parameters that a stratospheric aerosol SWCE intervention could have a significant impact on?

- *Parameters should be sought within each of the four categories of Radiative, Geophysical, Geochemical and Ecological.*
- *Spatial scales from the local to the global, and temporal scales from daily to centennial should be considered within each category.*
- *Patterns of natural climate variability (see Box 2.1.2.4) should be reflected within the identified parameters.*
- *A focus on unintended consequences must be included. Example questions include:*
    - o *How would various stratospheric aerosol interventions affect stratospheric chemistry, ozone, and ground-level ultraviolet radiation?*
    - o *What are the surface chemical and ecological impacts of precipitated aerosol particles?*
    - o *What deleterious climate changes might be expected? (for example, regional drying—see Box 2.1.2.5)*
    - o *How would natural and agricultural ecosystems be affected by changes in the total amount or spectral and angular distributions of sunlight? Such studies should include consideration of the combined affects of higher carbon dioxide (e.g., plant fertilization, ocean acidification), as well as the altered temperature, precipitation and sunlight.*

(2) What are the response functions for these important climate parameters, particularly defined in terms to the control variables for stratospheric aerosol interventions?

- *This will require clear definitions and articulation of the uncertainty associated with each of these response functions and sensitivity analyses defining the spectrum of potential climatic outcomes for given interventions.*

(3) What are the risk sensitivities of societal and ecological systems to these important climate parameters?

- *Beyond linear response, potential instabilities and non-linearity (in both climatic and dependent socio-political systems) should be examined rigorously.*

Analyses of socio-political systems and their sensitivities to changes in various climate parameters must play a central role in both the identification of the most important climate parameters to evaluate (Question 1) and any in risk-assessment methodology (Question 3.) For example, the sensitivity of the global agricultural system to changes in precipitation, temperature and ecological changes must be a central consideration in any evaluation of climate engineering. As no human systems exist in globally-averaged conditions, the spatial, temporal and climate parameter sensitivities of these systems must prominently guide this stream of research, together with the Ecological parameters. As such, social science considerations must be integrated into this research stream. The research agenda presented here focuses only on the engineering and technical aspects of these questions.

*Phase I* research efforts for all three questions core research questions for the CS stream would first draw upon existing climate system understanding and research efforts. Initial investigations would include significantly expanded computational modeling of stratospheric aerosol interventions. Box 2.1.2.2 discusses some limitations of the exploratory simulations to date, including particularly the limited incorporation of climate feedbacks and lack of investigation of impacts on natural patterns of climate variability. Greater realism is clearly required in these simulations (*e.g.* full spectral radiative effects of stratospheric aerosols, higher spatial resolution, dynamic ocean, sea ice and ecological models, etc.). These efforts will need to (minimally) include Atmospheric General Circulation Models (AGCMs), Ocean General Circulation Models (OGCMs), Sea Ice and Glacial dynamics models, and ecological models. Models that accurately reproduce natural patterns of climate variability (such as ENSO and NAO) should be used to investigate their sensitivity to SWCE (*see Box 2.1.2.4.*) The ensemble of results from simulations of the same (or very similar) scenarios on many independently validated and credible models would provide the most reliable and credible estimation of global and regional climatic responses.

A comparative modeling effort similar to that undertaken for IPCC AR4 would significantly improve our understanding. While we have not made a detailed estimate of the scale truly needed to sufficiently map climate response to SWCE interventions, a considerable increase in climate modeling efforts would certainly be required. As in the development of deployment technologies, primary effort should be directed towards understanding sulfate aerosol interventions, with secondary exploratory efforts dedicated to potential engineered aerosol alternatives.

*Phase I* laboratory experiments, as well as *Phase I* (non-intervention) field experiments for atmospheric chemical transport and interactions (such as those discussed in Section 3.1.1), could also support and enhance these modeling efforts. For ecological parameters, laboratory investigations can provide significant insights into how a wide range of biological organisms and ecological systems respond to changes in temperature, precipitation patterns, the ratio of direct to diffuse sunlight, and even storm patterns. Ongoing long-term studies of ecological responses to GHG-induced climatic change provide a baseline against which such laboratory investigations can be compared. Targeted interactions between the modeling and experimental communities must be facilitated—particularly for atmospheric chemical transport and interaction models and for ecological models.

Investigations of the climatologic and ecological responses to natural and unintentional anthropogenic events (both past, on-going, and future) that transiently alter radiative forcing (*e.g.* volcanoes, reduction of aerosol emissions during the Beijing Olympics or immediately post-9/11, etc.) would be another area of *Phase I* research. Much of this case study work is already being undertaken (*for example, see Box 2.1.2.3* and *Box 2.1.2.5*.) This could be complemented with a comparative examination of such cases to map the space of natural and unintentional anthropogenic analogues to stratospheric aerosol interventions. Moreover, this research avenue could substantially benefit from the international climate science community coordinating, in advance, preparations to observe future natural and anthropogenic events in greater detail. Evaluation of current readiness and coordination was beyond the scope of this report, but would clearly be a near-term, low-cost research objective.

It is the CS stream of research that would require the most expansive *Phase II* experiments. As discussed in Sections 3.1.1 and 3.2, low-magnitude, short-term field experiments will likely be sufficient to validate aerosol deployment systems, and to observe the temporal and spatial distribution of various aerosol materials in the stratosphere. However, the wide range of temporal and spatial delays and nonlinearities in the climate parameter response functions, along with the natural variability of these parameters, severely limit the utility of short-term tests for many parameters.

Of the four categories enumerated above, radiative climate parameters will be the most immediately and directly affected by a stratospheric aerosol intervention. Properly designed field-experiments of limited duration and scale could significantly enhance our empirical understanding of the radiative forcing properties of stratospheric aerosols, which would simultaneously contribute to the goals of the E stream. Such experiments could also reduce the uncertainty in current anthropogenic aerosol forcing levels, leading to better understanding of ongoing climate change. Similarly, iterative feedback between modeling efforts and experimental design would enhance the general quality of climate models—particularly those related to radiative flux.

However, as examined further in Section 3.2, basic issues with signal-to-noise and natural variability will fundamentally limit our ability to measure the steady-state response functions for many climate parameters. In some cases, it may take several years or decades of global, steady-state intervention (*Phases III-IV, see Sections 2.4 and 2.5*) to observe attributable parameter responses—and by that time, the intervention may have already committed such climate parameters to a much greater delayed response. Tools and experiments for maximizing our measurement capability are discussed in Sections 3.1.3 and 3.2 respectively.

### *3.1.3 CM Stream Research: Climate Monitoring Capabilities*

If field tests (*Phase II*) or deployment and maintenance (*Phases III* to *V*) of a SWCE system are undertaken, it will be essential to carefully measure climate impacts. Validation of climate science and modeling predictions, identification of unanticipated climatic impacts, and evolving understanding of societal and ecological risks will all critically depend on the timeliness and quality of available data for many climate parameters. Moreover, for many climate parameters, it would not be sufficient to have new monitoring systems in place only at the start of field tests or an intervention. To accurately observe the climatic parameter impacts, sufficient historic baseline data must be acquired before any interventions or field tests. The definition of 'sufficient' will vary widely between climate parameters, based on the natural variability and response time scales of each parameter—but years or decades may be required in some cases (such as Ecological parameters.) Particular attention should thus be paid to the early development and deployment of suitable monitoring capabilities, well before interventions are considered.

The first stage of the CM stream would build upon the effort described in the Climate Science stream to identify those important climate parameters in the Radiative, Geophysical, Geochemical, and Ecological categories. After identifying the parameters, existing monitoring tools, scientific networks, and historic datasets need to be assessed. In some cases, existing national and international research programs will already have in place valuable infrastructure, and the SWCE research process should be strongly integrated with existing programs. However, these available climate monitoring resources need to be evaluated against the (currently poorly defined) requirements of a dynamic control-system for any climate engineering system (*see Section 2.5.1.*) An iterative process of monitoring capacity and control-system design would be required—and, as above, we emphasize that a focus on considering the possibility of unintended climate parameter impacts is of paramount importance in this process.

A full review of existing global climate monitoring capabilities for any of the four identified categories of climate parameters is beyond the scope of this study. We provide an example of the type of evaluation that is recommended in Box 3.1.3.1, and define the key research questions for this stream generally as:

(1) What monitoring capabilities are required to confidently detect and assess the most important climatic impacts of a stratospheric aerosol SWCE intervention?

- *Here, capabilities include not only the physical data collection tools and infrastructure, but also the scientific networks, and data management data analysis capabilities. The requisite spatial and temporal scales, coverage, and degrees of accuracy and precision also need to be defined.*

(2) What monitoring capabilities presently exist to meet these requirements, and what new capabilities are needed? On what timescales can these tools be developed and deployed?

(3) How far in advance of an SWCE intervention (or field test) do these monitoring capacities need to be operational to provide the necessary calibration/background data?

Once the necessary climate monitoring capabilities are identified, the development and deployment of suitable monitoring tools falls entirely within *Phase I* (non-intervention) research. Moreover, each new monitoring capability can have direct benefits for ongoing climate change research, as well as in providing valuable new data and insights for the CS research efforts.

**Box 3.1.3.1 Monitoring the Earth's Radiative Balance**

As discussed in Box 2.1.2.1, the balance of incoming and outgoing radiation determines the Earth's average temperature. A first-order estimate of stratospheric aerosol impact on this radiative balance can be determined from the radiative properties and stratospheric lifetimes of the aerosol particles themselves. However, the net impacts of a stratospheric aerosol intervention on the Earth's radiative budget can be significantly amplified or attenuated through the climate system's radiative feedbacks. For example, atmospheric water vapor and clouds can significantly modify both the shortwave and longwave radiation fluxes,[91] and the formation of both in the atmosphere are in turn coupled to incident shortwave solar radiation and atmospheric aerosol concentrations. The snow and ice radiative feedback mechanisms discussed in Box 2.1.2.2 are also examples. These feedbacks governing the Earth's net radiation balance represent some of the largest sources of both uncertainty and sensitivity within the climate system.

Before field testing or deploying any SWCE, it will be critical to have in place a monitoring system that can observe both the solar irradiance (incident and reflected) and the radiance emitted by the Earth to space. Past or current systems such as ERBE and CERES may fulfill part of this need, but the monitoring requirements for SWCE will likely be beyond currently deployed capabilities and new systems will be required.

The tradeoffs between the characteristics of any new system and what is adequate to enhance confidence in SWCE interventions can only be assessed after detailed modeling and monitoring system design. However, one limit is a system that would provide the complete spectral signature of the radiative forcing of the Earth's climate, yielding significant insight into the response of the climate system to both increasing GHGs (*i.e.* ongoing climatic change) and any attempted SWCE. Two capabilities of such a monitoring system would be particularly valuable. First, solar radiation reflected from the Earth-atmosphere system back to space constitutes a powerful and highly variable forcing of the climate system through changes in snow cover, sea ice, land use, aerosol and cloud properties. Absolute spatially and spectrally resolved observations of the near ultraviolet, visible, and near infrared solar radiation incident upon and reflected by the Earth system could be important for the early observation of sensitive radiative feedback responses to any SWCE intervention.

Second, the system might measure the Earth's absolute spectrally resolved radiance in the infrared with high accuracy, likely by downward-directed spectrometers in Earth's orbit. These data could be used to both directly measure the IR forcing of the atmosphere of GHGs (important in its own right) and indirectly to evaluate the response of other atmospheric variables through the spectrally resolved signal of the outgoing IR radiance. In particular, large differences among climate model projections of temperature, water vapor and cloud distributions are correlated to different predictions for outgoing IR spectral signatures. The observed IR radiation signatures (with characteristics in frequency, spatial distribution and time) would thus provide new calibration metrics for existing climate models. Presently, longwave feedbacks for various climate models' realizations have a spread of ~25% between models,[92] and differences in radiative algorithms produce an even wider range of radiative forcings for prescribed GHG increases.[93] As such, these calibration signatures would help to significantly improve the quality of current climate models and thus understanding of the climate system.

These capabilities could both be incorporated into a new class of dedicated climate satellites. For any SWCE system, the number of climate satellites necessary to provide sufficient spatial and temporal resolution for the requisite multivariate control-system (*see Section 2.5.1*) should be carefully evaluated.

---

[91] Wetherald-1988; Held-2000; Colman-2003.

[92] Bony-2006.

[93] Collins-2006.

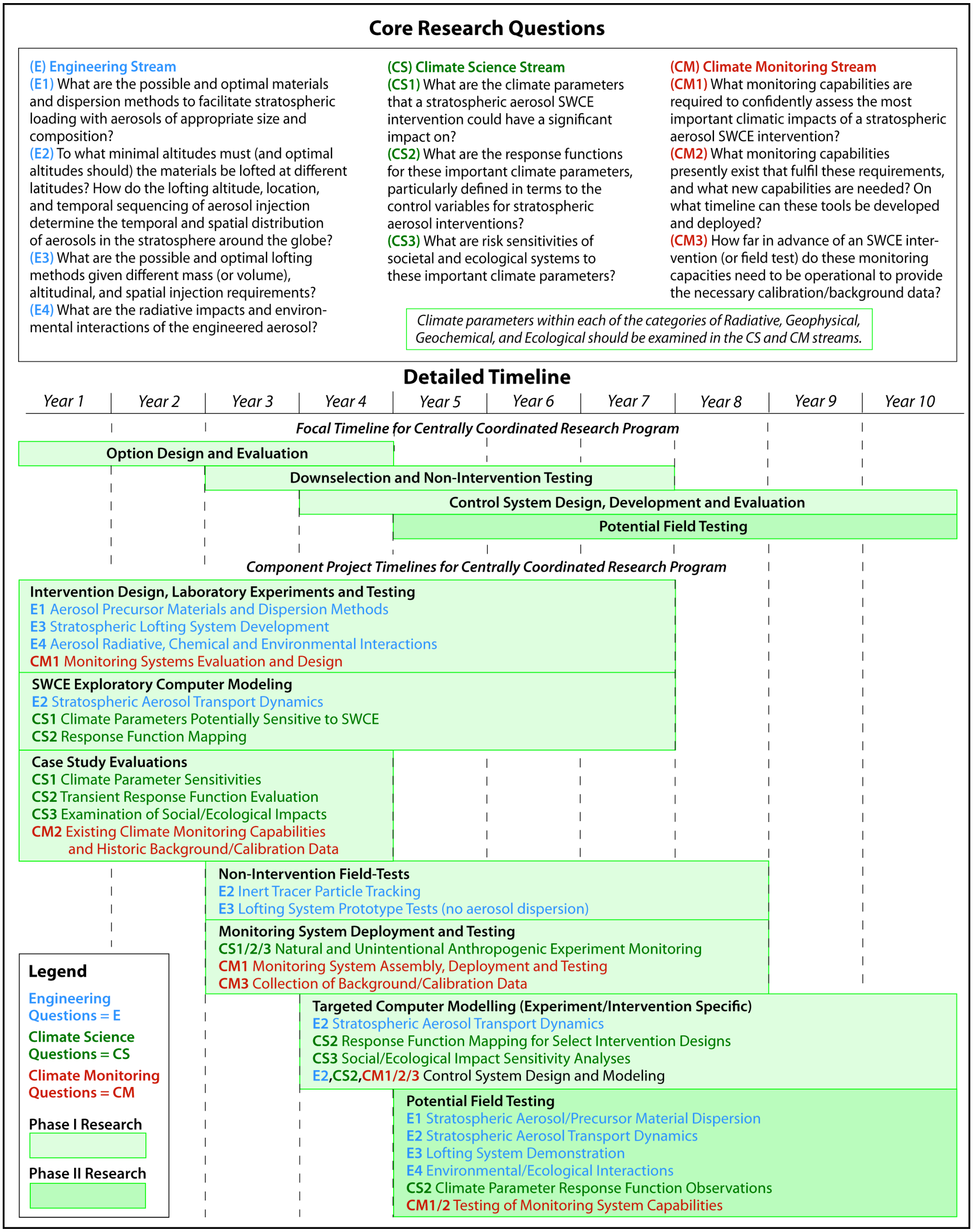

***Figure 5*** Outline of a comprehensive research program for stratospheric aerosol SWCE

Core research questions (top) and rough timeline (bottom) for a decade-long comprehensive research program to maximally reduce uncertainties associated with stratospheric aerosol SWCE.

### *3.2 The Value and Limitations of Field Testing*

Phase II field-testing involving stratospheric aerosol deployment could—if properly designed—benefit our understanding in each of the three research streams discussed in the previous section. With appropriate experimental design, even very short-term (<<1yr) and/or very low-level (<<0.1TgS, or equivalent) field-experiments could simultaneously provide demonstration of stratospheric lofting and aerosol dispersal mechanisms, observations of stratospheric aerosol transport and chemical interactions, and the opportunity to measure immediate radiative effects of stratospheric aerosols under clear-sky conditions. Such low-level, minimal climate impact experiments would be valuable for the iterative development of a complete stratospheric aerosol SWCE system. However there are important limits to what such subscale experiments could conclusively demonstrate.

Natural variability, along with complex feedbacks and delays in the temporal and spatial responses of the climate system, place fundamental limits on what could be determined from any short-term and/or low-level stratospheric aerosol deployment experiments. To exemplify this issue, we present simple calculations that place a lower bound on the field-test duration and scale required to conclusively demonstrate short-term impact on the surface temperature ($T_S$.) The methodology is applicable to any climate parameter, although the simplifying assumptions will be more or less accurate in different cases.

The following simple calculations assume that an optimized surface temperature monitoring system is already deployed and that sufficient background data from the system has already been collected. Moreover, we ignore both the temporal delays in global temperature response (due to the oceans' thermal inertia) and any ongoing additional perturbations to the global temperature (for instance, from GHGs or large climatic changes, such as the disappearing summer arctic sea ice.) The factors will only increase the timescales and magnitudes of field-test required to collect conclusive data.

Under these idealized conditions, we asked the question: how soon after the introduction of stratospheric aerosols would the effects on surface temperature ($T_S$) be conclusively detectable (i.e. rise above the natural variability) in the measured data? This question is fundamentally one of signal-to-noise ratio (*SNR*) and is crucial for the design of both field experiments (Phase II) and the ramp-up of a full intervention system (Phase III.) Moreover, the same calculations also provide an estimate of the duration of pre-intervention baseline data required to assess any impact on a given parameter.

Natural year-to-year fluctuations in global mean summer temperature are roughly Gaussian, with a standard deviation of $\sigma \sim 0.1$ºC. The globally-averaged radiative forcing ($F$) is related to temperature change by $\Delta T_S = F/\Lambda$, where $\Lambda \sim 2$ W/m$^2$•ºC. From these basic data (and assuming no year-to-year correlation of global temperatures) the *SNR* for a global experiment can be calculated through the relationship:

$$SNR = F\sqrt{N}/\Lambda\sigma$$

where $N$ is the number of successive time periods (in this case years) measured. This same calculation can be applied to regional-scale field-experiments. In that case, however, because heat is routinely carried in and out of regions by atmospheric and oceanic circulation (in contrast to the global case, for which net energy is conserved), both $\sigma$ and $\Lambda$ are larger.

For example, the standard deviation for the regional mean summer temperature is σ ~ 0.3ºC for the tropics, and σ ~ 3ºC for mid-latitudes, and for both cases, the local Λ is larger by perhaps a factor of 10 than the global case (but poorly known.)

The negative radiative forcing $F$ intended by SWCE might be a few $W/m^2$. While one would like *SNR* to be as large as possible, *SNR* values greater than 3 is necessary for any experiment to provide conclusive results. This appears feasible in a year for a full-scale global experiment, but not for a regional experiment. Indeed, a one-year global field-test involving a constant negative radiative forcing of 0.5 $W/m^2$ (estimated to yield of $\Delta T_S = 0.25$ºC) could generate a detectable effect. However, Crutzen's (2006)[94] estimate of 1 TgS inducing 0.75 $W/m^2$ implies a lofting sufficient to maintain a stratospheric loading of ~0.7 TgS for one year (equating to an steady-state injection rate of about 1 tonne per minute) would be required to achieve this radiative impact.

The *SNR* of an experiment might be improved substantially by appropriate modulation of the forcing with time and use of phase-sensitive detection, similar in principle to ``lock-in" signal detection. A more detailed calculation, including the measured spectrum of natural temperature fluctuations, shows that a perturbation $\Delta T_S = 0.1$ºC (estimated to require a stratospheric loading of ~0.2 TgS) modulated with a period of 6 years could reach *SNR* values of 3 in 25 years.

The implications of these *SNR* calculations for surface temperature are: 1) Any conclusive field-test (Phase II) will require a non-trivial stratospheric aerosol loading to be maintained for several years; 2) a global trial would become conclusive in a shorter duration than any regional experiment; 3) among regional experiments, a tropical experiment would become conclusive in a shorter period of time, simply due to the small natural fluctuations in tropical climate; and 4) modulation of the forcing could allow for shorter and/or smaller-magnitude experiments.

More extensive calculations associated with a comprehensive experimental design are clearly necessary to draw conclusions firmer than those presented here. Moreover, these conclusions apply only to the surface air temperature, and then only to prompt response. As discussed in Section 3.1.2, the climate parameters of importance will also span a wide range of physical and temporal scales. Particularly for climate parameters in the ecological category, significant temporal and spatial response delays will considerably complicate the basic *SNR* framework employed here.

---

94 Crutzen-2006.

# 4 Implications for the Socio-Political Dimensions of Climate Engineering

This study has focused only on the question of what technical research agenda could be pursued to maximally reduce technical uncertainty in climate engineering responses to climate emergencies. Socio-political factors and challenges were explicitly sidelined in order to facilitate an unconstrained assessment of the technical aspects. However, socio-political issues clearly cannot be ignored in—and in fact should be central to—any decision making process about climate engineering research and interventions. (*Note that the relation of climate engineering to mitigation efforts is addressed in the Prelude, and thus not readdressed here.*)

Our technical study raises some important socio-political questions that decision makers and social scientists need to considered carefully.

We briefly list those questions here without offering detailed analyses or solutions, and highlight issues where we believe detailed understanding of technical issues is particularly important for ensuring sound socio-political decisions.

An obvious question raised by this technical evaluation is: *Who decides?* This question applies to a wide range of specific issues, including:

(1) *Who decides* what climate engineering research should be pursued? (particularly important for field testing, where simultaneous climate engineering field tests would significantly complicate data analyses and increase risks)

(2) *Who decides* whether or which climate engineering alternative should be deployed?

(3) *Who decides* what the "optimal" target climate should be, both globally and regionally?

Our evaluation confirms that it is economically and technically possible for each stage of climate engineering research and deployment to be undertaken unilaterally (*see Sections 2.4, 2.5 and 3.1 respectively*.)[95] However, due to the highly international nature of current climate science research, independent national efforts over the next decade would be very unlikely to reduce technical uncertainty as much as a concerted international effort. Moreover, any field-testing (*Phase II*) components of any national programs would run the risk of interacting with similar testing of other national climate engineering research programs, thus interfering with ongoing international climate system research efforts. However, these facts do not preclude unilateral national research efforts or deployment. Similarly, the risks associated with countermeasures (*see Section 2.5.2*) do not preclude their unilateral use.

Optimally, international coordination or governance would guide all research (including placing limits on such research) and any deployment. However the structure associated with such an international system is obviously non-trivial. Existing international bodies such as the IPCC, UNFCCC, WCRP and IGBP[96] all provide models for international coordination of research and decision making on climatic issues, and each could play a central role in any climate engineering research program. Moreover, security and arms treaty bodies, as well as international medical organizations, provide examples of international coordination where the possibility of unilateral action (or lack thereof) poses significant near-term global risks.

---

[95] The possibility of unilateral climate engineering deployment has been discussed previously by: Schelling-1996; Barrett-2008.

[96] Intergovernmental Panel on Climate Change (IPCC), United Nations Framework on Climate Change (UNFCC), World Climate Research Program (WCRP) and International Geosphere-Biosphere Project (IGBP).

When considering international coordination or governance, the technical issues of response time scales must be a foremost consideration—particularly for governing full-scale deployment. As discussed in Section 2.5.1, the development of a dynamic multivariate control-system should be central to any deployed climate engineering intervention. Especially in the early stages of an intervention (*Phase III*) the management of such a control-system is almost certain to require rapid ($\ll$1yr) decisions about changes in some intervention variables. However, such rapid decision timeframes could be a major challenge for many existing international scientific and governance bodies—and the potential for unilateral disengagement and counter-engineering only exacerbates this rapid coordination challenge. Design of international processes, which include accounting for various possible domestic reactions and timeframes, should thus be a central focus of socio-political research on climate engineering.

Finally, if any climate engineering system is deployed, the attribution of resulting climate impacts could create a multiplicity of international challenges. In cases where climatic impacts can clearly be attributed, "winners and losers" resulting from those impacts might be identifiable, leading to issues of appropriate compensation.

However, as discussed in Sections 2.3 and 3.1.2, inherent temporal delays and natural variability in climate parameters will inevitably leave large uncertainties in our understanding of many climate response functions, making confident impact attribution impossible in many cases. Design of international processes for handling climate engineering should also thus consider methods for evaluating climate engineering impacts and arbitrating potential disputes that could arise.

Many other socio-political dimensions extending well beyond those considered here and in the Prelude to this report must be carefully evaluated. The *Phases* of climate engineering development and deployment described in Sections 2.4 and 2.5 provide a basic framework of technical decision points that can clarify the relevant socio-political dimensions at each point. Discussion of the cultural, ethical, legal, political and economic implications of climate engineering should not be avoided, because they will have a dominant influence on any decisions regarding even climate engineering research in the near future.

# Appendix

## Preliminary Evaluation of Stratospheric Lofting Methods

We examined a number of the most prominent lofting mechanisms proposed for placing aerosol materials (or their precursors) in the stratosphere. The issues examined are far from a comprehensive survey. Instead, we only provide a sample of the issues that require further investigation for each alternative, and even advance current thinking on some alternatives with rough calculations and estimates. This section demonstrates clearly the preliminary state of understanding of several of these frequently proposed methods. The example evaluations provided here are in no way a substitute for proper engineering design studies of these options.

### *Aircraft*

Lofting megaton quantities into the stratosphere requires heavy lift aircraft that can fly at these altitudes. As noted in Box 3.1.1.1, lofting to 20 km might be sufficient for deployment of stratospheric aerosols in the equatorial region—however, only detailed scientific investigations of atmospheric aerosol transport will be able to address this question.

Presently there are no aircraft designed specifically for this purpose. Close analogs for considering aircraft lofting potential could be the subsonic WB-57 or supersonic XB-70 (~23km ceiling, 250 ton max takeoff weight), or the more recent Theseus or White Knight Two (WK2.) WK2 is designed for rapid sorties above ~15 km with a payload estimated at around 10,000 kg. With some reengineering, a scaled and unmanned version of the WK2 craft might provide the capability to repetitively loft significant mass to ~20 km.

Assuming a nominal ~$10^9$ kg/yr injection rate[97] and a 10,000 kg lofting capacity for a specially designed aircraft, it would require ~100,000 sorties to be flown each year (or ~300 sorties per day.) With each craft assumed capable of two sorties per day, this would require a fleet of 150 aircraft. Conservatively estimating costs for a specially designed aircraft of up to $200M per aircraft, along with reasonable annual capital and O&M cost estimates (15%/yr capital and 5% per year O&M), the required aircraft fleet costs are roughly estimated to be ~$6B/yr. Further ~10,000 kg-fuel/sortie and $2/kg fuel-cost, yields fuel costs of another $2B/yr, bringing the total costs for the nominal ~$10^9$ kg/yr injection rate to ~$8B/yr. This corresponds to a cost of $8/kg, roughly an order of magnitude higher than current commercial airfreight rates. These costs do not include aerosols or dispersal equipment, and depend on the assumption that aerosols can be delivered just above the tropical tropopause—if substantially higher delivery is required, aircraft costs would go up dramatically.

### *Guns*

The option of using naval guns to fire shells containing the aerosol materials into the stratosphere was examined by the 1992 NAS report.[98] The summit of the naval artillery art was achieved in the Iowa class battleships, whose 50×16" guns[99] fired a 2,700 lb (~1,200 kg) shell with a muzzle velocity of 2500 ft/sec (762 m/sec) at a firing rate of two shells per minute.

---

[97] See Endnote 70 for estimates of the necessary stratospheric sulfate loading.

[98] NAS-1992, page 452.

[99] The first number is the barrel length in units of the caliber.

In the absence of an atmosphere, such a shell fired upwards would reach an altitude of 29.6 km. Air drag slows the shells by ~10%,[100] reducing the maximal attainable altitude to ~24 km.

This may be adequate for injection into the upwelling lower equatorial stratosphere, but would lead to short residence times in the downwelling polar stratosphere.

By comparison, the main gun on the M1 tank has a muzzle velocity up to 1.7 km/sec, illustrating what can be achieved with conventional chemical propulsion. This higher muzzle velocity clearly allows for substantially higher altitude lofting, but with tradeoffs in terms of lofted mass. For example, the 10 kg M829A3 round has a muzzle velocity of 1.555 km/sec, and if scaled to 406 mm, it could reach an altitude of 63 km with a total shell mass of 387 kg—of which 300 kg might be payload. Optimal design of a gun for stratospheric injection to 30–50 km requires engineering tradeoffs among these factors, but these altitudes are clearly achievable.

Davis guns are (in principle) recoilless; if their barrel is rifled they are also known as recoilless rifles. The absence of recoil is achieved with a barrel open at both ends, with a wad of soft material expelled out the back to take up the recoil of the projectile. Because of their small recoil, Davis guns are capable of firing massive projectiles without enormously robust and expensive mounts, but because of the recoil mass they are energetically inefficient and unlikely to be able to loft material into the stratosphere.

With the demonstrated firing rate of two rounds per minute for 16" naval guns, carrying a 1000 kg payload, a nominal ~$10^9$ kg/yr injection rate[101] would require three guns firing at their maximum rate (two shells per minute) year-round. Tank guns have a firing rate roughly ten times higher, but it is unclear how this would scale to 16" caliber, at which the scaled tank round would carry 1/3 the payload of the naval gun—given these parameters, only three tank guns would be required. The optimal gun would likely lie somewhere between these limits, depending on the desired altitude of injection.

Barrels of tank guns must also be replaced every few hundred rounds because of erosion. This might be less frequent at the lower muzzle velocity required here, however barrel replacement clearly needs to be considered. The chief consequence of barrel erosion is reduced accuracy, which is not a significant issue for stratospheric aerosol dispersion, so it is likely a safe assumption that each barrel is capable of firing 1000 rounds. A nominal ~$10^9$ kg/yr system with ~333 kg payloads per round would fire about 3,000,000 rounds/yr from a total of three tank guns, consuming ~3000 barrels/yr at this rate (or roughly ~10 barrels/day.)

The cost of the shells—not including the aerosol or precursor payload—might be up to ~$10,000 per shell. This is a very conservative guess as to the cost of these very simple shells—by comparison, the technically sophisticated JDAM guidance package is estimated to cost $20,000 per item. Mass production could reduce the unit cost far below that typical of low production-run peacetime military systems.

---

100 For this estimate, we assume for the shells a drag coefficient $C_D = 0.2$ and a ballistic coefficient of $BC = 944$ gm/cm$^2$, and approximate the atmospheric scale height is small compared to the altitude reached. The equation used is:

$$\frac{v_f}{v_i} = \exp\left[-\frac{C_D}{2}\left(\frac{1000\,gm/cm^2}{BC}\right)\right] \approx 0.9$$

101 See Endnote 70 for estimates of the necessary stratospheric sulfate loading.

This nominal cost corresponds to a lofting cost of \$30/kg, or \$30 billion for $10^9$ kg. More massive shells with parameters closer to those of the extant 16” naval guns would likely have a lower cost per unit mass lofted. The guns, barrels, and other components are likely to be a small fraction of the cost of the shells because of the economies of using them continually over a long period. Unlike military systems, no elaborate turret capable of aiming accurately over wide angles would be required—only a very limited range of vertical trajectories is required for stratospheric material lofting.

### *Rockets*

The required velocities 1–1.5 km/sec are a fraction of the exhaust velocity of solid rocket fuels. The payload mass fraction in this case would be between 0.6 and 0.7 of the launch weight.[102] This is greater than for high velocity tank rounds, in which the propellant mass exceeds the projectile mass, but probably less than that of naval gun rounds. Air drag is less important for rockets than for guns because rockets may be slenderer than gun-launched projectiles, and because the burnout velocity of a rocket is only achieved above the denser parts of the atmosphere.

Rockets could also provide additional advantages over guns, such as a milder launch environment, permitting the payload to be carried in a thin-walled vessel rather than a massive artillery shell, and the absence of a massive breach and barrel to contain the confined burning propellant. We did not make a quantitative estimate, but believe the cost of rocket lofting would also be less than that of guns.

### *Chimneys*

It may be possible to inject gases into the stratosphere through a tube running from the Earth’s surface to the stratosphere and suspended from a balloon (resembling a chimney.) The buoyancy of the balloon must be sufficient to support the full weight of the tube. It must also prevent tropospheric winds from turning the tube horizontal, thereby pulling its upper end below the required injection altitude.

The materials considered have been high molecular weight polymers such as Spectra, liquid crystal polymers such as Vectran, and aramids such as Kevlar. These materials all have uniaxial (aligned fiber) strength ≈ 30 KBar, Young’s modulus ~1 Mbar and density 1–1.5 gm/cm$^3$.[103] If the fibers are distributed orthogonally in a thin sheet its biaxial strength and modulus along the axes may approach half the uniaxial value, but it may be very weak in diagonal tension in which fibers can slide over one another. If the fibers are isotropically distributed very few will be aligned in any direction and the material’s tensile strength will depend on their resistance to sliding, not on their uniaxial tensile strength. The loads on the tube will be predominantly along its length, so its fibers may be oriented in that manner. However, it is unclear what tensile strength to assume for a balloon fabric in isotropic plane tension. For the tube, a tensile strength (*S*) of 10 Kbar may be reasonable, but for the balloon this could likely be much smaller.

---

102 Calculated using $1-\exp\{-v_b / v_e\} \approx 0.6–0.7$, where $v_b$ is the rocket velocity at fuel burnout (exhaustion).

103 Fette-2004.

Using standard methods,[104] we have also made a rough estimate of the tube diameter required to accommodate a nominal $10^{10}$ kg/yr (~300 kg/sec) flow through the tube,[105] assuming a driving pressure roughly comparable to ambient, as will be driven by buoyancy if the molecular weight of the gas is a fraction of that of air. One method of satisfying this condition would be to dilute a hydride gas precursor (*see Box 2.3.1.2*) two- or three-fold with hydrogen. The result is a radius ($r_t$) at the upper end of the tube, assuming a stratospheric pressure of 30 mbar (about 30 km altitude), to deliver 300 kg/sec ($5\times10^7$ $m^3$/day) of $r \approx 2$ meters. The lower end may be several times narrower because of the higher local pressure and consequent gas density.

Finally, any tube must pass through the troposphere into the stratosphere, and will thus occasionally encounter the jet stream. The aerodynamic load for a length $L$ immersed in a jet stream of speed $v = 50$ m/sec is:

$$F = C_D\, L\, v^2\, r_t\, \rho_a \sim 5 \times 10^{11}\ dyne$$

where we have taken $L = 1$ km as the jet stream depth, $C_D = 1$ and $\rho_a = 1\times10^{-3}$ gm/cm3 for the upper troposphere. It may be possible to reduce this load by a factor 2 or 3 if the tube is aerodynamically shaped with a "weather vane" to turn it into the wind. In addition, we have made the very conservative assumption that the tube has a constant diameter. In fact, the portion at jet stream altitudes may be a few times narrower because of the higher density there (we have tacitly assumed the gas in the tube to be in pressure equilibrium with the ambient air), reducing the aerodynamic load in proportion. The cross-section required to bear this load in tension is 50 $C_D/S$ $cm^2$. The weight of a tube of length $L_t = 50$ km is then:

$$W = g \times M = \left(\frac{g\rho_t L_t C_D}{S}\right) \cdot 4 \times 10^{11} dynes$$

Of course, the previous calculation is not self-consistent. The tube must bear its own weight as well as that of any aerodynamic load. We could solve the self-consistent equation, but instead make the following qualitative points:

1. The balloon must support a load ~ $10^{12}$ dynes (the weight of ~1000 tonnes.)
2. It is essential that the along-axis tensile strength of the tube material be on the order of 10 Kbar.
3. Minimizing lateral wind resistance by aerodynamic shaping and optimal orientation of the tube has large benefits.

Such chimney designs may be feasible, but they involve significant technical risks in material and aerodynamic performance over current levels. Though this section (and the evaluation of needed balloon lofting capacity in the following section on balloons) considers only the deployment of a single chimney to loft all the required materials, the same qualitative challenges (though with some quantitative differences) would apply to a configuration involving multiple chimneys at distributed locations lofting scaled down volumes of materials.

---

[104] Menon-2005.

[105] See Endnote 70 for estimates of the necessary stratospheric sulfate loading. These calculations allow for an order of magnitude higher mass flow rate than the nominal $10^9$ kg of aerosol materials required, allowing for significant dilution of the aerosol materials or precursors in the chimney.

### *Balloons*

As noted above, for the chimney method to work, stratospheric balloons must be deployed to support the tube(s.) Alternatively, balloons might also be used as lofting mechanisms in their own right. Both uses are considered here.

For a balloon to support a chimney, daily expenditure of lift gas or ballast needs to be avoided. The solution is an overpressure balloon, whose volume is essentially independent of its temperature. The concept is old, but its realization has depended on the development of better materials, which may now be available.

To provide $10^{12}$ dynes of lift at the 30 mbar level requires a volume of about $2\times10^{13}$ $cm^3$, or a radius ($r_b$) of ~170 m. The overpressure ($\Delta P$) it must support is a fraction $f_{var}$ (the fractional diurnal temperature variation) of ambient, or ~$10^4$ dyne/$cm^2$. The required wall thickness is therefore:

$$\Delta r_b \geq \frac{f_{\mathrm{var}}\Delta P}{2S} r_b \approx \frac{0.01}{S} cm$$

where $S$ is the tensile strength of the material. The ratio of the weight $W_{skin}$ of its skin to its buoyant lift ($B$) is then:

$$\frac{W_{skin}}{B} = \frac{3}{2}\frac{f_{\mathrm{var}}\Delta P}{10^{10} S}\frac{\rho_{skin}}{\rho_a\left(1-\mu_b/\mu_a\right)} \approx 0.05/S$$

where we have taken a temperature of 250ºK, $\rho_{skin}$ of ~1.5 gm/$cm^3$, $f_{var}$ of 0.3, and $H_2$ as the filling gas for the a balloon (with molecular weight $\mu_b$), and $\mu_a$ is the molecular weight of the displaced air.

The importance of the material strength is evident. For example, Mylar has an ultimate tensile strength ($S$) of 1.5 Kbar,[106] which gives $W_{skin}$/B of ~0.3, so the requirement to contain the overpressure of large temperature swings would exact a large price in a Mylar balloon's lifting capability. The materials discussed above for the tubes are much stronger, but their behavior when used to make membranes subject to isotropic tension must be studied before their suitability can be determined.

If balloons are used instead as individual lofting mechanisms, than long-residence time is not required. In this case overpressure is not required, and the buoyancy of a balloon in pressure equilibrium with the ambient air is:

$$B = g\left(M_a - M_b\right) = gM_a\left(a - \frac{\mu_b}{\mu_a}\frac{T_a}{T_b}\right)$$

where $M_a$, $\mu_a$ and $T_a$ are the mass, molecular weight and temperature of the displaced air and $M_b$, $\mu_b$ and $T_b$ are the mass, molecular weight and temperature of the gas filling the balloon. If $T_a = T_b$ than the buoyant life ($B$) is independent of altitude (and, equivalently, independent of atmospheric pressure) and such a balloon has no equilibrium height. If $B$ exceeds the load it will rise indefinitely, until (if open at the bottom) it spills lifting gas, or (if closed) it bursts from internal overpressure once the skin expands to its maximum volume. If $B$ is less than the load it sinks to the surface of the Earth.

---

106 Koehler-1995.

In practice, the altitude of a pressure-equilibrium balloon—such as those used to loft scientific payloads to the stratosphere—is controlled by dumping ballast. If the filling gas were always in thermal equilibrium with the air it would remain at a constant altitude indefinitely, once enough ballast had been dumped (or gas spilled) that the lift equals the load. But because of the diurnal variation in solar heating of the skin (and advective heat transport to the filling gas) the ratio of $T_a$ to $T_b$ varies, and ballast or gas must be expended daily.

As a result, pressure-equilibrium balloons have flight durations on the order of ten days, except during polar summer and winter ("midnight Sun" or "noontime night") when there is no diurnal Solar heating cycle.

Somewhat longer durations may be obtained if: (1) they are made of material that is less absorbing of Solar near-infrared radiation than polyethylene (thereby reducing the magnitude of their diurnal temperature swings); (2) they are baffled inside to reduce transport of heat from the skin; or (3) they are aluminized to reflect sunlight.

These balloons must be in pressure equilibrium with the ambient air because they are made of very weak material (typically 0.8 mil polyethylene, with a tensile strength of ~ 300 bars and even lower yield threshold, so that a 100 m radius balloon begins plastic flow at an overpressure of < ~100 dyne/cm$^2$ (~$10^{-4}$ bar) or about $10^{-2}$ of a float pressure of 10 mbar at about 120000 ft.

Such balloons are very cheap and light, and have been used to loft scientific payloads for many years. For simple stratospheric lofting (and not sustained tube suspension), these balloons may be satisfactory. Gaseous material may either be mixed with hydrogen or helium as the lifting gas, or (if liquid or solid) may be suspended from the balloon, as scientific payloads are. Volatile materials (such as the precursor hydrides proposed in Box 2.3.1.2) could be carried as gases to take advantage of their buoyancy, which partially offsets their weight. This could also avoid the need to lift the parasitic weight of a cryogenic or pressurized container.

Balloon delivery of materials has been considered and rejected by Rasch *et al*. (2008)[107] on the grounds that the number of balloons is excessive and that the large number of expended balloons falling to the Earth would pose an unacceptable risk to the environment. Balloons that failed to vent or burst might also pose a risk to aviation upon their unpredictable descent. These objections are difficult to evaluate, and would require further engineering study. A typical scientific balloon operation may cost several hundred thousand dollars, and lofts a payload of order a ton, suggesting a cost per unit mass roughly 1 to 10 times that of artillery lofting—though dedicated research and mass production could again lower these costs. The launch of such a balloon is a tricky operation that depends on favorable weather (low wind) at the launch site.

---

[107] Rasch-2008a.